\documentclass[fleqn,usenatbib]{mnras}

\usepackage{newtxtext,newtxmath}

\usepackage[T1]{fontenc}

\DeclareRobustCommand{\VAN}[3]{#2}
\let\VANthebibliography\thebibliography
\def\thebibliography{\DeclareRobustCommand{\VAN}[3]{##3}\VANthebibliography}

\usepackage{graphicx}	
\usepackage{amsmath}	
\usepackage{physics}
\usepackage{caption}
\usepackage{subcaption}
\usepackage[normalem]{ulem}

\title[Implicit dust dynamics]{An implicit algorithm for simulating the dynamics of small dust grains with smoothed particle hydrodynamics}

\author[D. Elsender \& M. R. Bate]{
Daniel Elsender\thanks{E-mail: de296@exeter.ac.uk (DE)}
and Matthew R. Bate
\\
Department of Physics and Astronomy, University of Exeter, Stocker Road, Exeter EX4 4QL, UK}

\date{Accepted XXX. Received YYY; in original form ZZZ}

\pubyear{2023}

\begin{document}
\label{firstpage}
\pagerange{\pageref{firstpage}--\pageref{lastpage}}
\maketitle

\begin{abstract}
We present an implicit method for solving the diffusion equation for the evolution of the dust fraction in the terminal velocity approximation using dust-as-mixture smoothed particle hydrodynamics (SPH). The numerical scheme involves casting the dust diffusion equation into implicit form, rearranging into its resolvent cubic equation and solving analytically. This method is relevant for small grains that are tightly coupled to the gas, such as sub-micron dust grains in the interstellar medium or millimetre-sized dust grains in protoplanetary discs. The method avoids problems with the variable used to evolve the dust fraction becoming negative when evolved explicitly and is fast and accurate, avoiding the need for dust stopping time limiters and significantly reducing computational expense. Whilst this method is an improvement over using the explicit terminal velocity approximation method, as with any dust-as-mixture method it still fails to give accurate solutions in the limit of large (weakly coupled) grains.
\end{abstract}




\section{Introduction}

Dust emission is the dominant signature in cold astrophysical observations. Small grains absorb ultraviolet/optical light emitted from stars and re-emit this energy in the infrared. Instruments such as the mid-infrared instrument (MIRI) onboard the James Webb Space Telescope (JWST) is primed to detect these emissions from dust. Infrared observations can be used in conjunction with radio observations, such as those using the Atacama Large (sub)Millimeter Array (ALMA), to explore the structures of protoplanetary discs. Within protoplanetary discs it is dust that provides, at least in part, the material from which planets form. In addition to this, dust grains collide with gas particles, altering  the dynamical evolution of both phases. Understanding the dynamics of dust in these environments is crucial for the interpretation of observations that rely upon assumptions about the dust-to-gas ratio, as dust structures can be vastly different to structures in the gas.

Smoothed particle hydrodynamics (SPH) is a popular numerical method for modelling astrophysical fluid dynamics.
Previously dust and gas mixtures have been modelled in SPH using separate particles for the dust and gas, originally by \citet{monaghan_sph_1995}, where the dust is coupled to the gas via drag. Such methods are sometimes referred to as two-fluid methods, but are less ambiguously referred to as dust-as-particles. This dust-as-particles method was applied in the context of protoplanetary discs by \citet{maddison_building_2003} and \citet{barriere-fouchet_dust_2005}. Subsequently, \citet{laibe_dusty_2012a,laibe_dusty_2012b} presented a new method for simulating dust-as-particles with SPH. Grid codes have also been adapted to model dust-as-particles. \citet{bai_particle-gas_2010} introduced a hybrid particle-gas scheme to simulate dust-as-particles in dusty gas mixtures with the grid code \textsc{athena}; they also present a fully implicit solver. The grid-code \textsc{fargo3d} uses an implicit numerical scheme to model multiple species of dust \citep{benitez-llambay_asymptotically_2019}. Implicit timestepping methods have also been employed in SPH dust-as-particle methods.  \citet{monaghan_implicit_1997} modified this method to treat the drag terms implicitly via pairwise interactions. More recently, \citet{laibe_dusty_2012b} introduced an implicit timestepping method for their dust-as-particle method based upon the pairwise interaction method of \citet{monaghan_implicit_1997}. \citet{loren-aguilar_two-fluid_2014, loren-aguilar_two-fluid_2015} developed a semi-implicit approach to dust-as-particle methods in SPH, although this scheme is limited to the use of global timesteps. \citet{stoyanovskaya_two-fluid_2018, stoyanovskaya_simulations_2020} presented an implicit dust-as-particles SPH method for both linear and non-linear drag regimes.

Modelling the dust as a set of particles in essence doubles the memory required to perform the calculation and usually significantly increases the time to completion compared to an equivalent calculation with only gas. In the past decade much effort has gone into understanding and modelling the dynamics of small dust grains, that are highly coupled with the gas phase. In a series of papers \citet{laibe_dusty_2014a,laibe_dusty_2014b, laibe_dust_2014c} discuss the limitations of the dust-as-particles method in the high drag (small grain) regime. To circumvent these issues they developed a dust-as-mixture method (sometimes called one-fluid) for modelling highly coupled dusty gas mixture by framing the equations in the barycentric reference frame of the fluid. \citet{price_small_grains_2015MNRAS.451..813P} presented a simplified version by adopting the terminal velocity approximation \citep{youdin2005streaming} and modelling the dust fraction evolution as a diffusion-like equation. The dust diffusion method is far less computationally expensive than an explicit dust-as-particles method in the limit of small grains. In addition to being deployed in SPH, the dust-as-mixture method for modelling small grain dynamics has been implemented in the adaptive mesh refinement code \textsc{RAMSES} \citep{lebreuilly_small_2019}.

The dust-as-mixture method of \citet{price_small_grains_2015MNRAS.451..813P} has been extended to model the dynamics of multiple small dust grain populations by \citet*{hutchison_multigrain_2018}. This dust-as-mixture method is an effective way to simulate the dynamics of small dust grains in the limit of the terminal velocity approximation. However, although the method enforces positivity on the dust fraction, $\epsilon$, (by evolving a variable $s= \sqrt{\epsilon \rho}$), where $\rho$ is the total mass density) it does not limit the dust fraction to less than unity. This issue was solved by \citet{ballabio_dust_fraction_2018} who use a different parameterisation of dust fraction in their evolution equations. Namely the dust fraction is defined as 
\begin{equation}
    \epsilon = \frac{s^2}{1+s^2},
    \label{eq:Ballabio}
\end{equation}
and it is the dust variable $s$ that is evolved. An alternative formulation using a dust variable defined by $\epsilon = \sin^2(s)$ was used by \citet{hutchison_presolar_2022}.  This formulation also enforces $\epsilon \in [0,1]$ and was found to produce more accurate solutions than equation \ref{eq:Ballabio}, but it is more computationally expensive due to the need to evaluate trigonometric functions.

For all of the above formulations, while the positivity of the dust fraction is guaranteed, there is nothing stopping the dust variable, $s$, itself from becoming negative. In the regions of simulations where there are steep dust fraction gradients, while enforcing exact dust mass conversation, the dust variable can undershoot zero and become negative. If it does so this can cause the dust fraction to become positive in these regions where, otherwise, there should be no dust. This consequence of the mathematics of the model can be thought of as dust leaking unphysically into these dust sparse regions. This can become a significant issue when modelling dust grains that are moderately coupled to the gas, for example, in protoplanetary discs where larger dust grains would be expected to settle to the midplane of the disc they can `leak' into or be retained in the upper, and lower, regions of the disc relative to the midplane.  This has been avoided in the past by sacrificing exact dust mass conservation and simply neglecting any change in the dust fraction or dust variable that would make it negative \citep[e.g., see section 4.4.2 of][]{price_small_grains_2015MNRAS.451..813P}.  The resulting dust mass conservation is better when using equation \ref{eq:Ballabio} or the $\epsilon = \sin^2(s)$ formulations than when using $s=\sqrt{\epsilon \rho}$ or evolving the dust fraction directly, but it is still not ideal.

An additional limitation of the dust-as-mixture method is that it requires a timestep criterion whereby timesteps become small in regions of low dust fraction, or where dust grains become less well coupled (i.e., large). To prevent calculations having prohibitively short timesteps, \citet{ballabio_dust_fraction_2018} introduced a mass flux limiter to stop rapid dust diffusion for particles in regions of high dust fraction gradients by artificially limiting the stopping time. The problem with this is that it unphysically modifies the drag acting on the dust during a simulation and, although it provides a numerical solution, the degree to which the solution resembles reality is unclear. 

In this paper, we present an implicit method for solving the dust diffusion equation of \citet{price_small_grains_2015MNRAS.451..813P}, using the dust variable defined by equation \ref{eq:Ballabio} \citep{ballabio_dust_fraction_2018}. This avoids the need for a dust timestep criterion and negates the need for a stopping time limiter.  In addition, it better handles the evolution of the dust variable in regions of large dust gradients due to the iterative nature of the algorithm. Depending on the dust grain sizes, the method also significantly speeds up the time to completion of three-dimensional simulations of protostellar collapse and protoplanetary disc evolution simulations, and yet yields equivalent results to those obtained using the fully explicit method.

Our aim with this work is to improve calculations in the limit of small, coupled grains.  Although we have applied this to the particle-as-mixture method that uses the terminal velocity approximation, a similar method could be derived to evolve the dust-as-mixture equations without the additional terminal velocity approximation. Applying such an implicit method to the full dust-as-mixture equations of \citet{laibe_dusty_2014a,laibe_dusty_2014b} would in principle allow grains with long stopping times to be more accurately modelled. This may be worth investigating in the future, but it is beyond the scope of this paper.  Furthermore, while the full set of dust-as-mixture equations can provide a better model for intermediate-sized grains than applying the terminal velocity approximation, the fundamental assumption of a particle-as-mixture method that the velocity field is single valued does eventually break down, as has been shown by \citet{laibe_dusty_2014b} and \citet{bate_dynamics_2017}.  Finally, we note that \citet{hutchison_dust_2016} used an implicit method to evolve the the differential velocity between gas and dust in full dust-as-mixture SPH method, but they still solved the dust diffusion equation explicitly.

\section{Method}

\subsection{The dust-as-mixture dust method}
\label{sec:exp_one_fluid}

In the derivation of the dust-as-mixture formulation \citep[][]{laibe_dusty_2014a}, the continuum equations are rewritten in the barycentric reference frame of the fluid. This involves replacing the velocities of the gas and dust phases with a single barycentric velocity of the mixture,
\begin{equation}
    \boldsymbol{v} = \frac{\rho_\mathrm{g} \boldsymbol{v}_\mathrm{g} + \rho_\mathrm{d} \boldsymbol{v}_\mathrm{d}}{\rho_\mathrm{g} + \rho_\mathrm{d}},
\end{equation}
where $\boldsymbol{v}$ is the velocity, $\rho$ is the density and the subscripts d and g denote the dust and gas quantities, respectively. The differential velocity between the two phases is defined as,
\begin{equation}
    \Delta \boldsymbol{v} = \boldsymbol{v}_\mathrm{d} - \boldsymbol{v}_\mathrm{g}.
\end{equation}
The total density of the mixture is,
\begin{equation}
    \rho = \rho_\mathrm{g} + \rho_\mathrm{d},
\end{equation}
and the dust fraction is defined as,
\begin{equation}
    \epsilon \equiv \rho_\mathrm{d}/\rho.
\end{equation}
The dust-as-mixture equations can then be written in the form
\begin{align}
    &\dv{\rho}{t} = - \rho \left( \div{\boldsymbol{v}}\right),\\
    &\dv{\epsilon}{t} = - \frac{1}{\rho} \div{\left[ \epsilon\left(1-\epsilon\right) \rho \Delta \boldsymbol{v} \right]}\label{eq:exponefluiddtep},\\
    &\dv{\boldsymbol{v}}{t} = - \frac{\grad{P}}{\rho} - \frac{1}{\rho} \div{\left[ \epsilon\left(1-\epsilon\right) \rho \Delta \boldsymbol{v}^2  \right]} + \boldsymbol{f},\\
    &\dv{\Delta \boldsymbol{v}}{t} = - \frac{\Delta \boldsymbol{v}}{t_\mathrm{s}} + \frac{\grad{P}}{\left(1-\epsilon \right)\rho} - \left( \Delta \boldsymbol{v} \cdot \grad{} \right) \boldsymbol{v}
     +\frac{1}{2} \div{\left[ \left(2\epsilon - 1\right)\Delta\boldsymbol{v} \Delta\boldsymbol{v} \right]},\\
    &\dv{u}{t} = - \frac{P}{\rho_\mathrm{g}}\div{\left( \boldsymbol{v} - \epsilon \Delta \boldsymbol{v} \right)}+ \epsilon \Delta\boldsymbol{v} \cdot \nabla u + \epsilon \frac{\Delta\boldsymbol{v}^2}{t_\mathrm{s}},
\end{align}
where $P$ is the gas pressure, $\boldsymbol{f}$ represents non-fluid forces acting on the mixture (e.g., gravity),  $u$ is the internal energy of the gas, and $t_\mathrm{s}$ is the stopping time of the dust. 

The stopping time when the dust is in the Epstein drag regime can be expressed as 
\begin{equation}
    t_\mathrm{s} = \frac{\hat{\rho}_\mathrm{s}r_\mathrm{s}}{\rho_\mathrm{g}v_\mathrm{th}},
\end{equation}
where $\hat{\rho}_\mathrm{s}$ is the intrinsic grain density, $r_\mathrm{s}$ is the grain radius, and the velocity of the gas molecules due to thermal motion is
\begin{equation}
    v_\mathrm{th} = \sqrt{\frac{8k_\mathrm{B}T}{\pi\mu m_\mathrm{H}}},
\end{equation}
where $k_\mathrm{B}$ is Boltzmann's constant, $T$ is the gas temperature, $\mu$ is the mean molecular weight of the gas, and $m_\mathrm{H}$ is the atomic mass of hydrogen.

This formalism can be simplified when the stopping time is small compared to the hydrodynamic time scale \citep{price_small_grains_2015MNRAS.451..813P}. In SPH terms this means $t_\mathrm{s} < h/c_\mathrm{s}$, where $h$ is the SPH particle smoothing (resolution) length and $c_\mathrm{s}$ is the sound speed. When this is satisfied, we are in the terminal velocity regime \citep{youdin2005streaming}, i.e. the relative velocity of the dust and gas reaches terminal velocity quickly due to the drag and pressure forces balancing. In this regime the dust-as-mixture equations can be simplified by ignoring the time dependence of the differential velocity, and any terms that are second order in $t_\mathrm{s}$. Then
\begin{equation}
    \Delta v = t_\mathrm{s}\frac{\grad{P}}{\rho_\mathrm{g}}.
\end{equation}
The continuum equations then become
\begin{align}
    &\dv{\rho}{t} = - \rho \left( \div{\boldsymbol{v}} \right), \label{eq:mass}\\
    &\dv{\epsilon}{t} = - \frac{1}{\rho}\div{\left( \epsilon t_\mathrm{s}\grad{P}\right)},\label{eq:dustfrac_evo}\\
    &\dv{v}{t} = - \frac{\grad{P}}{\rho} + \boldsymbol{f},\label{eq:dvel-contin}\\
    &\dv{u}{t} = - \frac{P}{\rho_\mathrm{g}}\left( \div{\boldsymbol{v}} \right) + \frac{\epsilon t_\mathrm{s}}{\rho_\mathrm{g}} \left( \grad{P} \cdot \grad{u} \right).\label{eq:energy}
\end{align}
The inclusion of equation \ref{eq:dustfrac_evo} means an additional time step criterion is required. \citet{price_small_grains_2015MNRAS.451..813P} provide such a criterion in the form
\begin{equation}
    \Delta t < C_\mathrm{0} \frac{h^2}{\epsilon c_\mathrm{s}^2 t_\mathrm{s}},
\end{equation}
where $C_\mathrm{0}$ is the Courant number. This implies that the time steps are constrained when the stopping times are long, i.e. when drag is low / dust grains are not well coupled to the gas. These equations can be discretised into the SPH form. Notably, the continuity and momentum equations have the same discretisation as the gas-only SPH equations. We must have new discretisations for the dust fraction, and the thermal energy evolution equations. 

To discretise the dust fraction evolution equation whilst maintaining positivity of the dust fraction, and limiting the fraction to below unity, thus prevents the dust fraction from becoming unphysical, we use the parameterisation of equation \ref{eq:Ballabio} introduced by \citet{ballabio_dust_fraction_2018} resulting in the dust evolution variable
\begin{equation}
    s = \sqrt{\frac{\epsilon}{1-\epsilon}}.
\end{equation}
We compute the time evolution of this variable 
\begin{equation}
    \dv{s}{t} = \frac{1}{2s\left(1-\epsilon\right)^2}\dv{\epsilon}{t},
    \label{eq:ddustvarepsilon}
\end{equation}
using a direct second derivative \citep{price_small_grains_2015MNRAS.451..813P}.  Substituting Eq. \ref{eq:dustfrac_evo} into Eq. \ref{eq:ddustvarepsilon} yields
\begin{equation}
    \dv{s}{t} = - \frac{1}{2\rho ( 1-\epsilon )^2} \{ \div{[ s(1-\epsilon)t_\mathrm{s}\grad{P} ]}+(1-\epsilon)t_\mathrm{s}\grad{P}\cdot\grad{s} \}.
\end{equation}
Using the implementation of \citet{ballabio_dust_fraction_2018}, the SPH discretisation of this equation is
\begin{align}
    \dv{s_i}{t} = - \frac{1}{2\rho_i (1-\epsilon_i)^2} \sum_j \left[ \frac{m_j s_j}{\rho_j}( D_i + D_j) ( P_i - P_j) \frac{\bar{F}_{ij}}{r_{ij}} \right],\label{eq:sphstdustvar}
\end{align}
where $D_i = t_{s,i}(1-\epsilon_i)$,  $\bar{F}_{ij} = \left[ F_{ij}(h_i) + F_{ij}(h_j) \right]/2$, $r_{ij} = | \boldsymbol{r}_{ij}|$, 
and $\vb{\hat{r}}_{ij}F_{ij}(h_i) = \grad_i W_{ij}(h_i)$, where $W_{ij}(h_i)$ is the SPH kernel. This work utilises the M$_{6}$ quintic spline kernel.

The energy of the system can be written as 
\begin{equation}
E = \sum_i m_i \left[ \frac{1}{2} v^2_i + (1 - \epsilon_i) u_i \right],
\end{equation}
which is conserved if
\begin{equation}
    \dv{E}{t} = \sum_i m_i \left[ v_i\dv{v_i}{t} + (1 - \epsilon_i) \dv{u_i}{t} - u_i \dv{\epsilon_i}{t} \right] = 0.
\end{equation}
Removing the non-dust terms, assuming they conserve energy, and substituting in our parameterisation for dust fraction gives
\begin{align}
    \eval{\dv{E}{t}}_\text{dust} &= \sum_i m_i (1 - \epsilon_i) \eval{\dv{u_i}{t}}_\text{dust} \nonumber \\ &\quad \quad  - \sum_i m_i u_i \frac{2 s_i}{\left(1 + s_i^2\right)^2} \dv{s_i}{t} = 0,
\end{align}
using Eq. \ref{eq:sphstdustvar} and rearranging  yields
\begin{equation}
\begin{split}
        \eval{\dv{u_i}{t}}_\text{dust} = - \frac{1}{2 (1 - \epsilon_i)} \sum_j & m_j\frac{s_i s_j}{\rho_j} \left( D_i + D_j \right) \\ &\left( P_i - P_j \right) \left( u_i - u_j \right) \frac{\bar{F}_{ij}}{r_{ij}}.
\end{split}
\end{equation}

When incorporated into the usual SPH equations for a gas, equation \ref{eq:mass} can be expressed as
\begin{equation}
    \dv{\rho_i}{t} = \sum_j m_j (\vb{v}_i - \vb{v}_j) \cdot \grad_i W_{ij}(h_i),
\end{equation}
although this equation is not explicitly solved as the density is computed by the usual weighted sum over neighbours, and equations \ref{eq:dvel-contin}, and \ref{eq:energy} become
\begin{align}
     \dv{\vb{v}_i}{t} &= - \sum_j m_j \left( \frac{P_i}{\rho_i^2 \Omega_i} \grad_i{W_{ij}(h_i)} + \frac{P_j}{\rho_j^2 \Omega_j} \grad_j{W_{ij}}(h_j) \right) \nonumber \\ &+ \boldsymbol{f} + \Pi_\text{AV},\\
     \dv{u_i}{t} &= \frac{P_i}{\Omega_i\rho_i^2} \sum_j m_j (\vb{v}_i - \vb{v}_j) \cdot \grad_i{W_{ij}(h_i)} + \left(\dv{u_i}{t}\right)^\text{diss}\nonumber\\
    &- \frac{1}{2(1-\epsilon_i)} \sum_j m_j \frac{s_i s_j}{\rho_i \rho_j} \left( D_i + D_j \right) \left( P_i - P_j\right)\left( u_i - u_j\right)\frac{\Bar{F}_{ij}}{r_{ij}},
\end{align}
where $\Omega_i$ is the term related to the smoothing length gradients defined as
\begin{equation}
    \Omega_i \equiv 1 - \pdv{h_i}{\rho_i} \sum_j m_j \pdv{W_{ij}(h_i)}{h_i}.
\end{equation}
The $\Pi_{AV}$ term provides shock capturing via artificial viscosity, which in the SPH code we used is formulated as
\begin{equation}
    \Pi_{ij} = - \sum_j \frac{m_j}{\rho_{ij}} \left[ \frac{q_{ij}^i}{\rho_i \Omega_i} \grad_i{W_{ij}(h_i)} + \frac{q_{ij}^j}{\rho_j \Omega_j} \grad_j{W_{ij}(h_j)} \right],
\end{equation}
where $q_{ij}$ is defined as 
\begin{equation}
    q^i_{ij} = 
    \begin{cases}
        - \frac{1}{2} \rho_i v_{\text{sig},i} \vb{v}_{ij} \cdot \vb{\hat{r}}_{ij}, & \vb{v}_{ij} \cdot \vb{\hat{r}}_{ij} < 0\\
        0, & \text{otherwise,}
    \end{cases}
\end{equation}
where $\rho_{ij} = (\rho_i + \rho_j)/2$, and the signal velocity $v_{\text{sig},i}$ is
\begin{equation}
    v_{\text{sig},i} = \alpha_i^{\text{AV}} c_{\text{s},i} + \beta^{\text{AV}} \left\vert \vb{v}_{ij} \cdot \vb{\hat{r}}_{ij} \right\vert,
\end{equation}
and $\alpha_i^\text{AV} \in [0,1]$ and $\beta^{\text{AV}}=2$. We use the method of \citet{Morris_Monaghan_1997} to evolve $\alpha_i^\text{AV}$.  Finally, the contribution of the artificial viscosity to the gas internal energy is
\begin{equation}
\left(\dv{u_i}{t}\right)^\text{diss} = - \frac{1}{\Omega_i} \sum_j \frac{m_j}{\rho_{ij}} v_{\text{sig},i} \frac{1}{2} (\vb{v}_{ij} \cdot \vb{\hat{r}}_{ij})^2 F_{ij}(h_i).
\end{equation}

\subsection{Implicit dust evolution algorithm}
\label{sec:imp_one_fluid}

Equation \ref{eq:sphstdustvar} is in the explicit form for the evolution of the dust variable, $s$.  Here we present a method for solving the dust diffusion equation implicitly using a backwards Euler method. This formalism uses Gauss-Seidel iterations to solve the implicit dust evolution equation. The idea of using this method to solve the dust evolution equation implicitly was inspired by \citet*{whitehouse_2005MNRAS.364.1367W}, who used this method to solve the diffusion equation for radiative transfer in the flux-limited diffusion approximation.  Using the implicit method detailed below allows us to use timesteps that are not constrained by the stopping time of the dust, instead using the hydrodynamical timestep and so typically speeding up calculations. 

To advance a time dependent variable, $A$, from time $t=n$ to $t=n+1$ the backwards Euler method can be stated as
\begin{equation}
	A_{i}^{(n+1)} = A_{i}^{(n)} + \mathrm{d}t \left( \dv{A_i}{t} \right)^{(n+1)}.
\end{equation}
This gives us the implicit equation for $A$. For interactions between particles $i$ and $j$, a Gauss-Seidel iterative method $A_i^{(n+1)}$ can be solved for by rearranging the implicit equation to the form 
\begin{equation}
	A_{i}^{(n+1)} = \frac{\displaystyle A_i^{(n)} - \mathrm{d}t\sum_j \sigma_{ij} \left( A_j^{(n+1)} \right)}{\displaystyle  1-\mathrm{d}t\sum_j \sigma_{ij}}, \label{eq:gauss-seidel}
\end{equation}
where $\sigma_{ij}$ contains quantities other than $A$. In this scheme $A_j^{(n+1)}$ starts off as $A_j^{(n)}$ and is updated to $A_j^{(n+1)}$ as soon as this value becomes available. Eq. \ref{eq:gauss-seidel} is iterated over until some specified convergence criterion is met.
Taking Eq. \ref{eq:sphstdustvar}, and putting it into backwards Euler form we obtain

\begin{equation}\label{eq:BEuler}
\begin{split}
    s_i^{(n+1)} = s_i - &\mathrm{d}t 
    \left \{ \frac{1+\left(s_i^{(n+1)}\right)^2}{2\rho_i}t_{\text{s}_i} \sum_j s_j^{(n+1)} L_{ij} +  \right. \\
    & \left. \frac{\left( 1+ \left(s_i^{(n+1)} \right)^2\right)^2}{2\rho_i} \sum_j \frac{t_{\text{s}_j}}{1+\left(s_j^{(n+1)}\right)^2} s_j^{(n+1)} L_{ij} 
    \right \},
\end{split}
\end{equation}

\noindent
where $L_{ij} \equiv \frac{m_j}{\rho_j}(P_i-P_j)\frac{\overline{F}_{ij}}{ r_{ij} }$. The implicit equation for $s_i^{(n+1)}$ can be rearranged into a quartic equation of the the form 
\begin{equation}
	a_4 \left(s_i^{(n+1)}\right)^4 + a_3 \left(s_i^{(n+1)}\right)^3 + a_2 \left(s_i^{(n+1)}\right)^2 + a_1 s_i^{(n+1)} + a_0 = 0, \label{eq:dustvarquartic}
\end{equation}
where
\begin{align*}
	a_4 &= \frac{\mathrm{d}t}{2\rho_i} \sum_j \frac{t_{\text{s}_j}}{1+\left(s_j^{(n+1)}\right)^2} s_j^{(n+1)} L_{ij}^{(n+1)},\\
	a_3 &= 0,\\
	a_2 &= \frac{\mathrm{d}t}{\rho_i} \sum_j \frac{t_{\text{s}_j}}{1+\left(s_j^{(n+1)}\right)^2} s_j^{(n+1)} L_{ij}^{(n+1)} + 
    \mathrm{d}t\frac{t_{\text{s}_i}}{2\rho_i} \sum_j s_j^{(n+1)} L_{ij}^{(n+1)},\\
	a_1 &= 1,\\
 \begin{split}
	a_0 &= \frac{\mathrm{d}t}{2\rho_i} \sum_j \frac{t_{\text{s}_j}}{1+\left(s_j^{(n+1)}\right)^2} s_j^{(n+1)} L_{ij}^{(n+1)} + \\ & \quad \mathrm{d}t\frac{t_{\text{s}_i}}{2\rho_i} \sum_j s_j^{(n+1)} L_{ij}^{(n+1)} - s_i^{(n)}.
 \end{split}
\end{align*}
A solution to Eq. \ref{eq:dustvarquartic} can be found with any method of solving quartic equations, e.g. numeric, however as $a_\mathrm{3} = 0$ this is a depressed quartic and we can solve the resolvent cubic (see Appendix \ref{App:qaurtic} for the analytic solution of the quartic). This analytic solution is preferred as it finds a solution more quickly. Iterating the solution to Eq. \ref{eq:dustvarquartic} over all SPH particles until convergence gives the values of the dust variable, $s$, at the next time step. Again, we note that $s_j^{(n+1)}$ starts off as $s_j^{(n)}$ until we have an updated value for $s_j$. 

\subsection{Convergence criterion}

The value of $s_i^{(n+1)}$ is iterated over until convergence within an accepted tolerance on the $m ^\text{th} $ iteration, i.e. iterations continue until 
\begin{equation}
    \left\vert \frac{s_i ^{n+1, m} - s_i ^{n+1, m-1}}{s_i ^{n+1, m}}\right\vert < \delta,
\end{equation}
where typically $\delta = 10^{-3}$ and $s_i ^{n+1, m}$
is the backward Euler form of $s_i^{n+1}$ (equation \ref{eq:BEuler}) evaluated at iteration $m$. If the method fails to converge, the timestep is split into two and the iterations begin again. We continue in this manner until convergence is reached, or until the original timestep has been split so many times that it is no longer computationally viable.

In the limit where $\epsilon \ge 2.5 \times 10^{-7}$ $(s > 0.0005)$ the equation is solved in this way. If the dust fraction drops below this value equation \ref{eq:dustvarquartic} is solved linearly (i.e. simply considering the last two terms on the left-hand side) as the leading order terms of the quartic become vanishingly small. We find no measurable loss of accuracy by solving the diffusion equation linearly in this limit.

\section{Numerical tests}

We have implemented both the implicit and explicit forms of the dust-as-mixture equations using the \citet{ballabio_dust_fraction_2018} dust parameterisation into the 3D SPH code \textsc{sphNG}. Originally \textsc{sphNG} was developed by \citet{benz_review_1990nmns.work..269B} and \citet{ benz_1990ApJ...348..647B} and subsequently modified significantly as described in \citet*{bate_sphng_1995MNRAS.277..362B}, \citet{whitehouse_2005MNRAS.364.1367W}, \citet{ price_2007MNRAS.374.1347P} and other papers. 

To test the dust-as-mixture dust fraction implementations, we used the \textsc{dustywave} and \textsc{dustshock} tests as outlined in \citet{laibe_dustybox_2011}. In these tests we adopt a stopping time defined as
\begin{align}\label{eq:ts_price}
    t_\mathrm{s} = \frac{\epsilon (1 - \epsilon) \rho}{K},
\end{align}
to be consistent with the literature and analytic solutions, where $K$ is the coefficient of drag between the gas and the dust. 

\subsection{Dustywave}
\label{sec:dustywave}

\begin{figure*}
    \centering
    \includegraphics[width=0.47\linewidth]{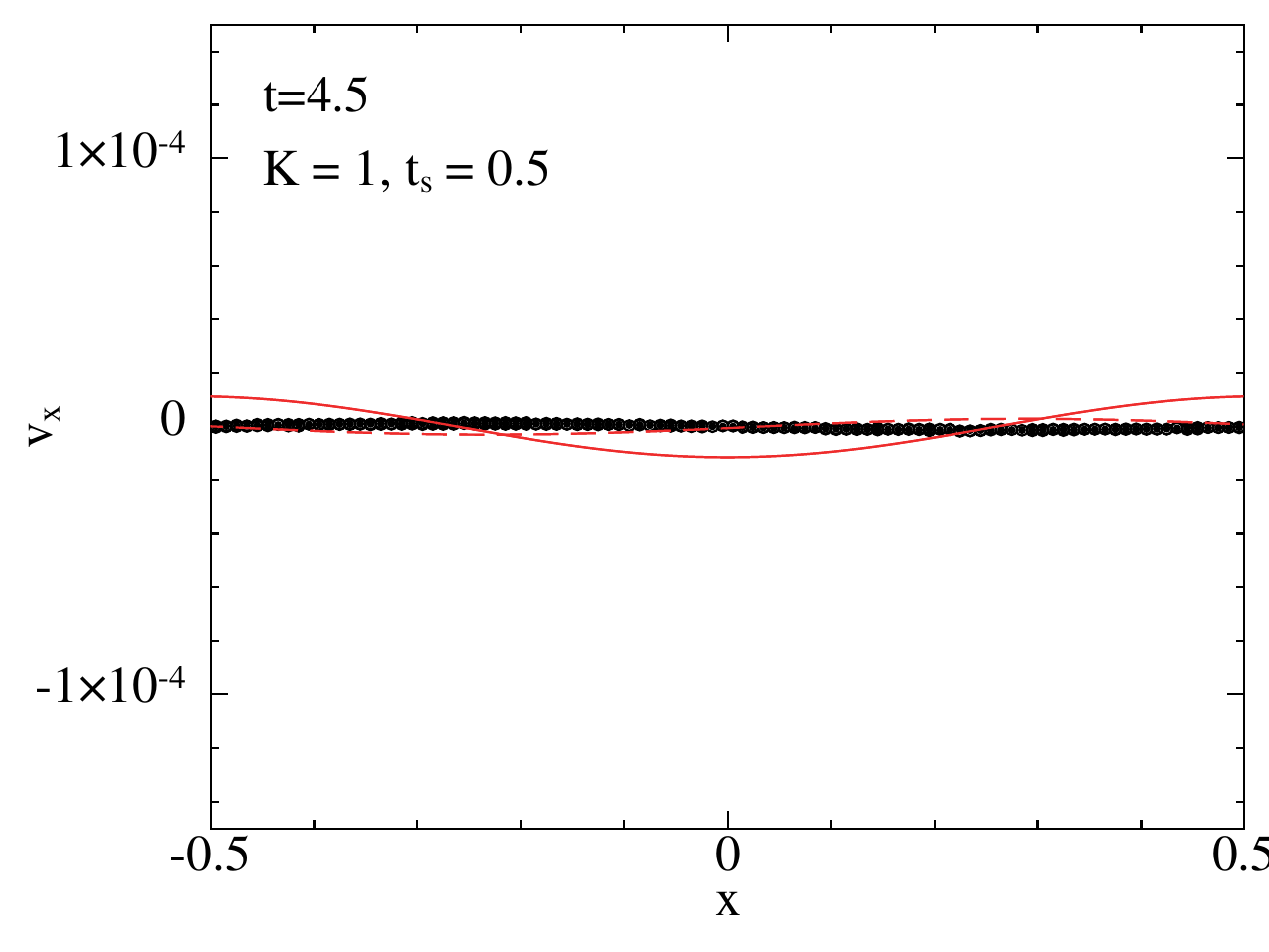}\hfill
    \includegraphics[width=0.47\linewidth]{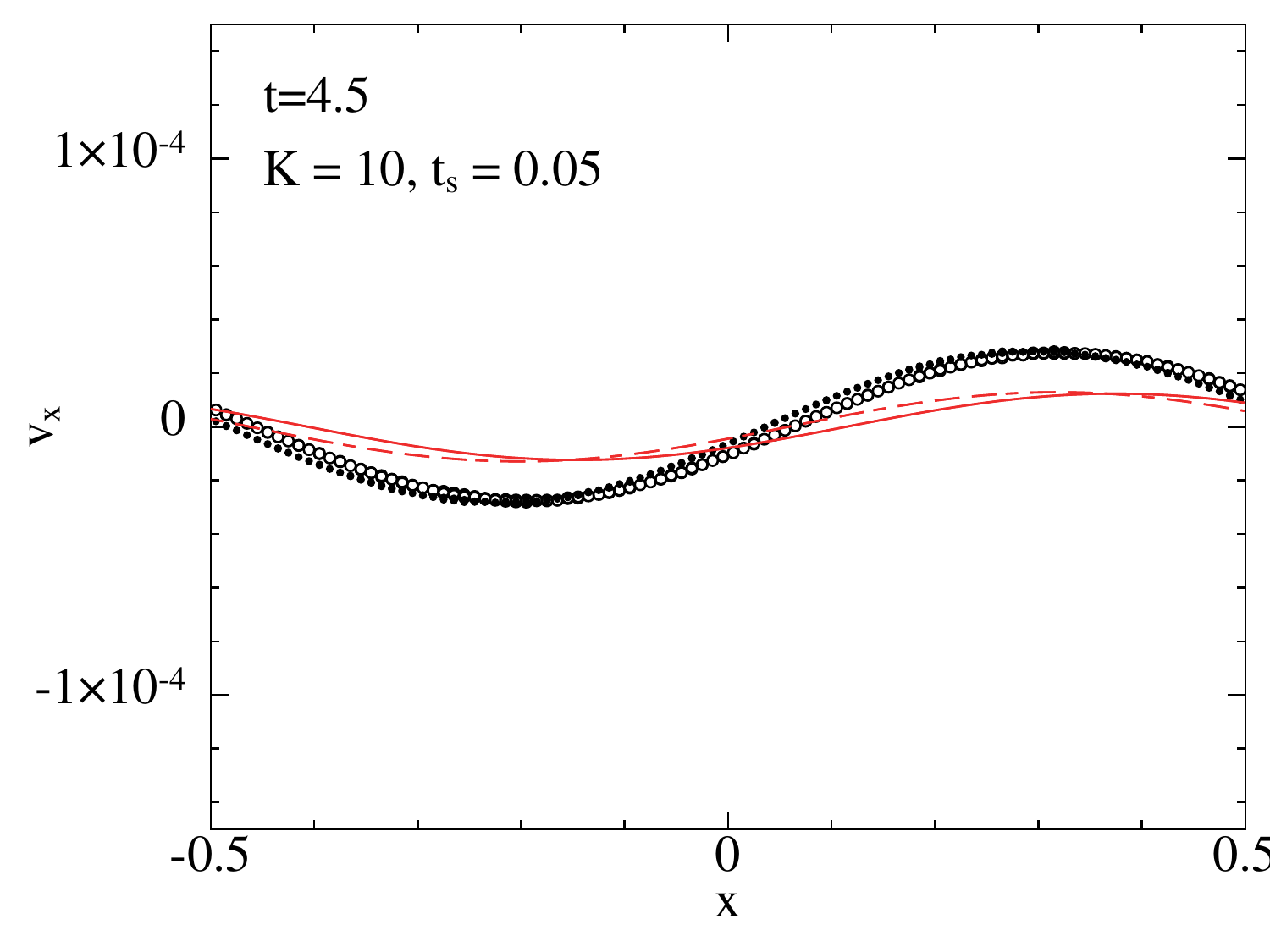}\hfill
    \includegraphics[width=0.47\linewidth]{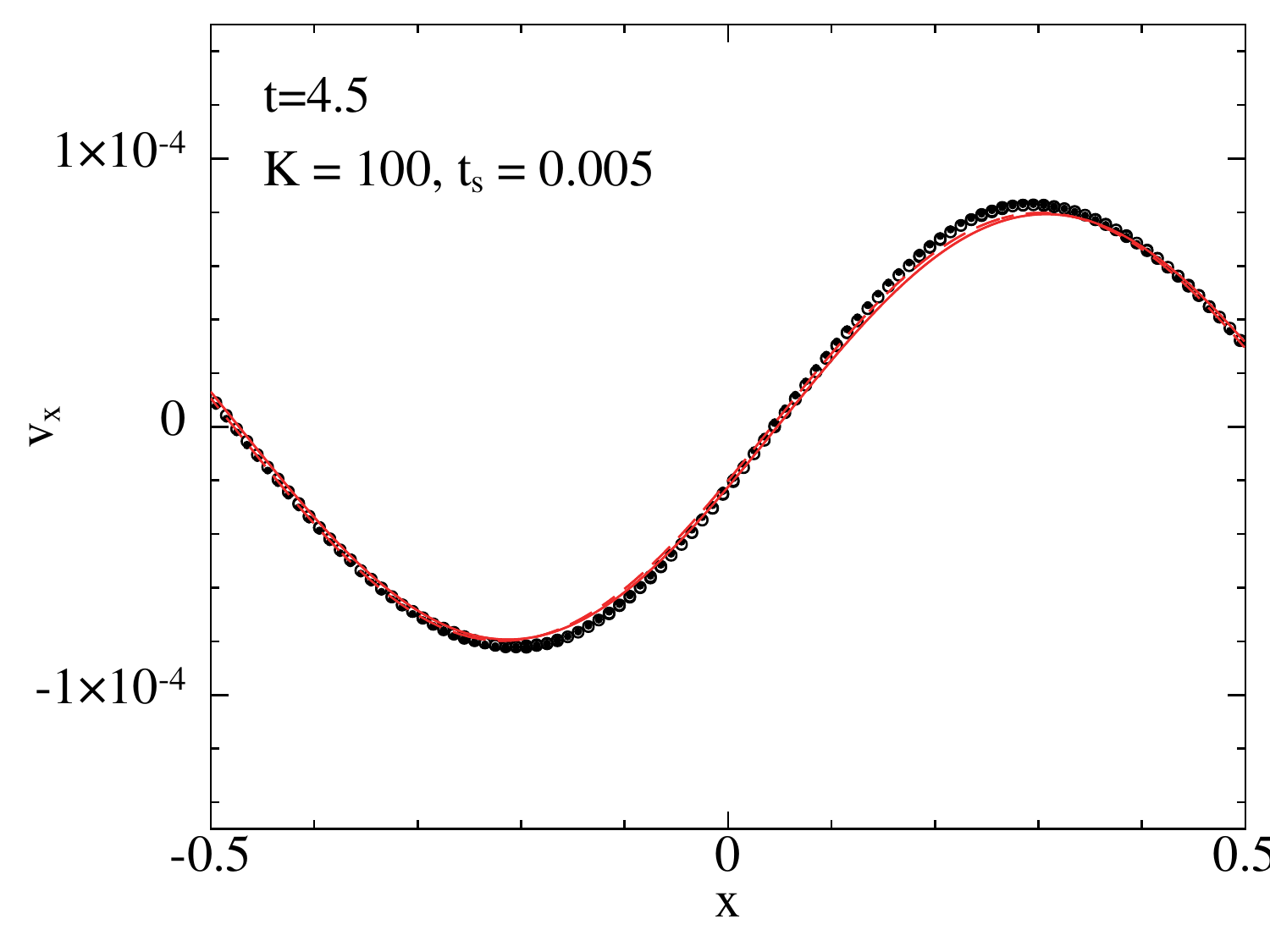}\hfill
    \includegraphics[width=0.47\linewidth]{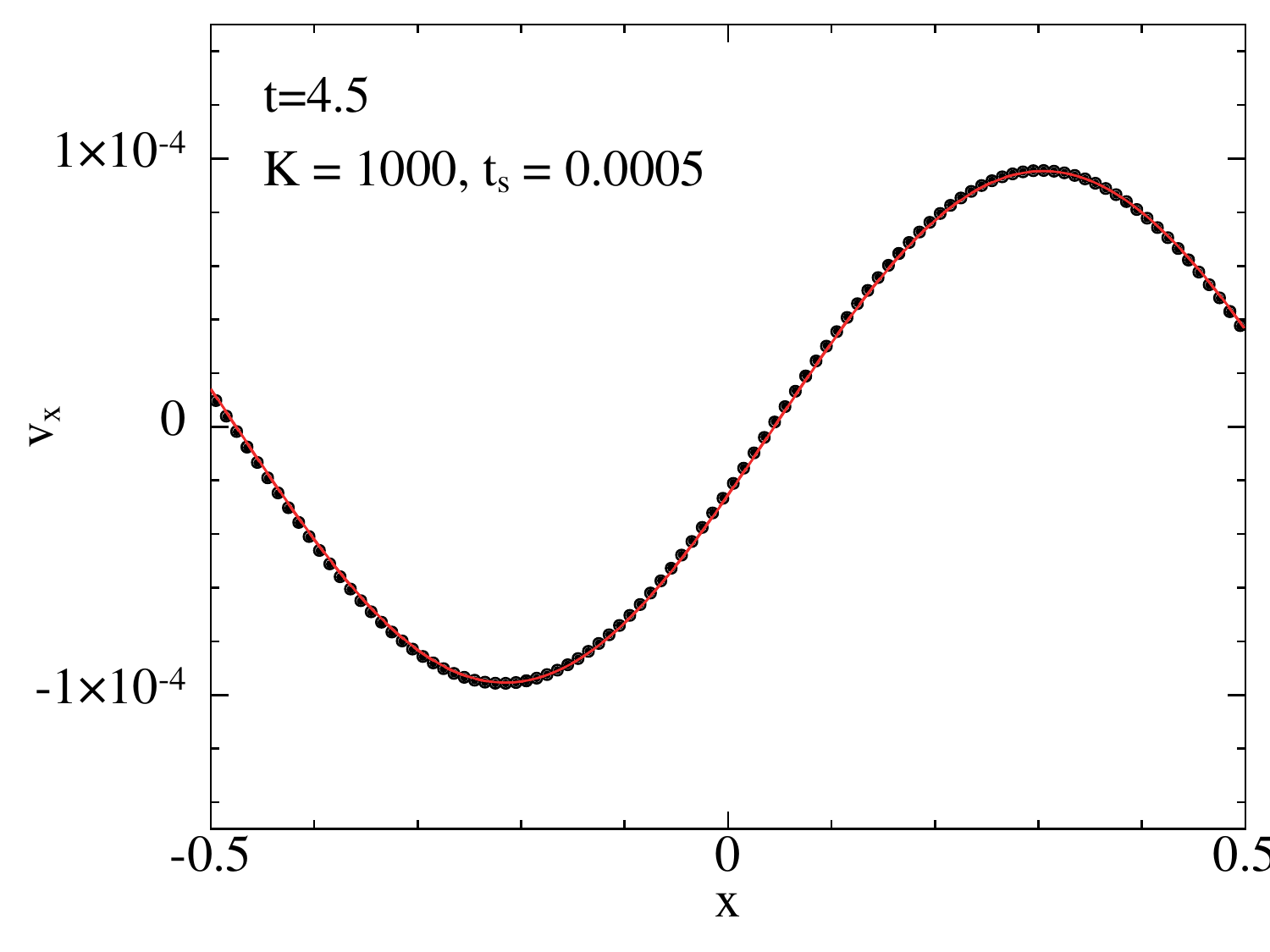}
    \caption{Results from the \textsc{dustywave} problem performed with our implicit method with a varying drag coefficient $K=1,\ 10,\ 100,\ 1000$ (from top-left to bottom-right) and a dust-to-gas ratio of 1:1. The analytic solution of the gas velocity is plotted in red as a solid line and the analytic solution of the dust velocity us plotted as a dashed line. The numerical solution of dust and gas velocities are plotted at time $t=4.5$ as open and filled circles respectively. As with an explicit method, the solution becomes inaccurate at lower drag ($K \lesssim 40$) due to the breakdown of the terminal velocity approximation.}
    \label{fig:dustywave}
\end{figure*}

The \textsc{dustywave} problem consists of a linear wave propagating through a dust-gas mixture. The problem was originally described for dust-as-particles by \citet{laibe_dustybox_2011}, then in dust-as-mixture by \citet{laibe_dust_2014c}. We set the problem up in 3D, but only propagate the wave in the $x$ direction. The problem is initialised with a mixture consisting of equal dust and gas ($\rho_\mathrm{d} = \rho_\mathrm{g} = 1$, i.e. $\rho_0 = 2$, $\epsilon = 0.5$) with a sinusoidal perturbation in both density and velocity of the form $v(x) = v_0\sin{(2 \pi x)}$ and $\rho(x) = \rho_0 \left[ 1 + \delta \rho_0 \sin{(2 \pi x)}  \right]$. The amplitude of the perturbation is set to $v_0 = \delta \rho_0 = 1 \times 10^{-4}$, with a corresponding perturbation in the energy. An adiabatic equation of state is used where $\gamma = 5/3$ and the sound speed is set to $c_\mathrm{s} = 1$. The domain is $x,\ y,\ z \in [-0.52,0.52],\ [-0.069,0.069],\ [-0.073,0.073]$ respectively, in which 3744 particles are set up in a closed packed lattice, giving a spacing of $\Delta p = 0.02$, to ensure particle stability throughout the calculation.

As noted by \citet{price_small_grains_2015MNRAS.451..813P}, there is an inconsistency between the analytic and numerical solution. In the analytic solution $\Delta \boldsymbol{v}$ is assumed to be zero\textbf{ at $t=0$}, however in the terminal velocity approximation $\Delta \boldsymbol{v} = t_\mathrm{s} \boldsymbol{f}$ is non-zero. This inconsistency causes a small phase difference between the analytic and numerical solution that gets larger as $t_\mathrm{s}$ gets larger, i.e. less drag / larger grains, and specifically as the terminal velocity approximation breaks down. 

The solution is plotted after $\approx 4.5$ wave periods in Fig. \ref{fig:dustywave} in which both gas and dust velocities are plotted in filled and open circles respectively. These values are reconstructed from the barycentric values for velocity as follows
\begin{align}
    \boldsymbol{v}_\mathrm{g} &= \boldsymbol{v} - \epsilon \Delta\boldsymbol{v},\\
    \boldsymbol{v}_\mathrm{d} &= \boldsymbol{v} + (1 - \epsilon) \Delta\boldsymbol{v}.
\end{align}
The results from our implicit method are consistent with those of \citet{price_small_grains_2015MNRAS.451..813P} for the explicit method implemented using a direct second derivative.
For both explicit and implicit methods, the solutions are very close to the analytic solutions in the regime where the terminal velocity approximation is valid ($K\gtrsim 40$). In lower drag regimes, the terminal velocity approximation gives a solution that is slightly out of phase or is over-dampened in the lowest drag case (see the two upper panels of Fig. \ref{fig:dustywave}).

\subsection{Dustyshock}
\label{sec:dustshock}

\begin{figure*}
    \centering
    \includegraphics[width=0.47\linewidth]{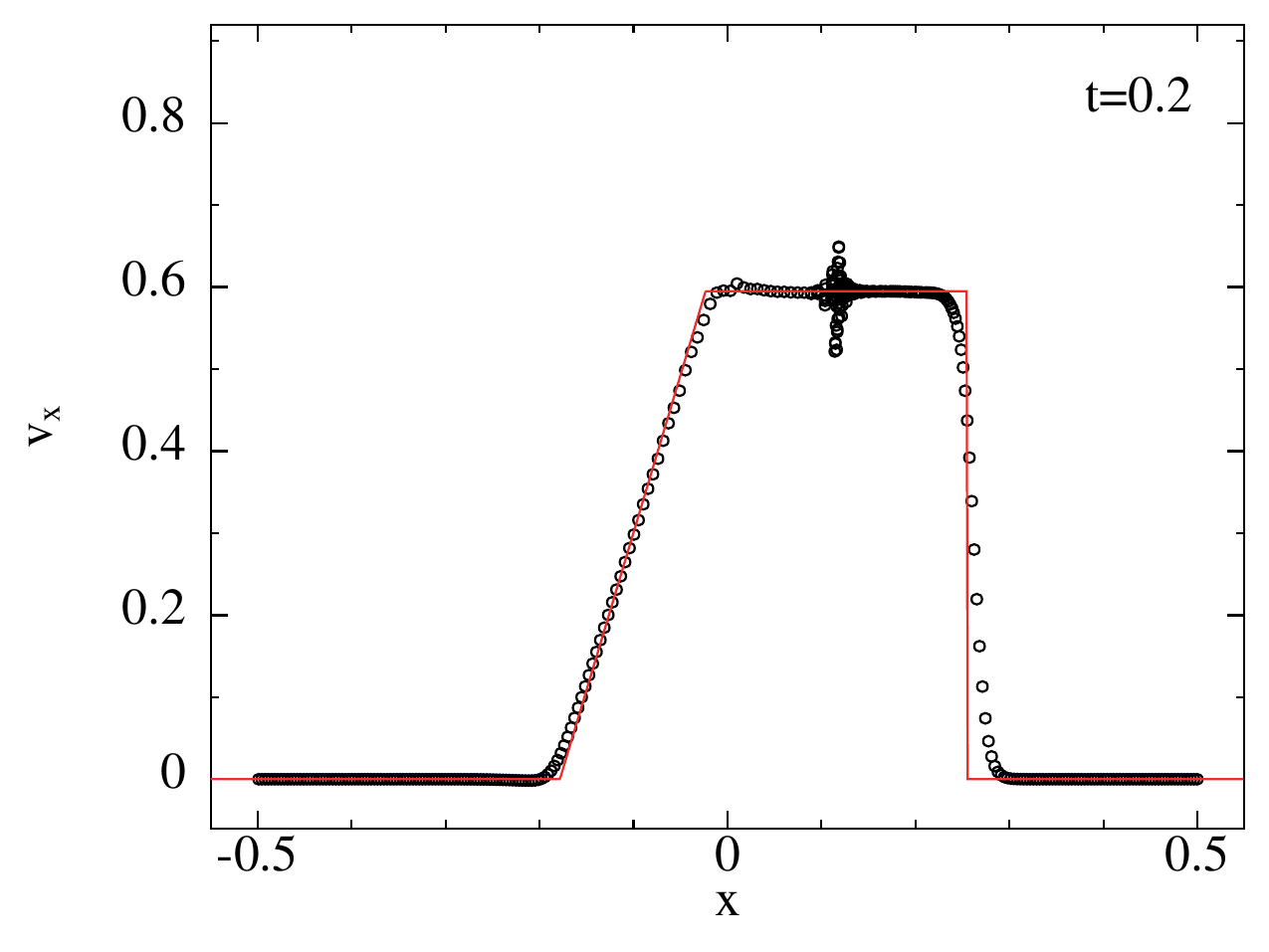}\hfill
    \includegraphics[width=0.47\linewidth]{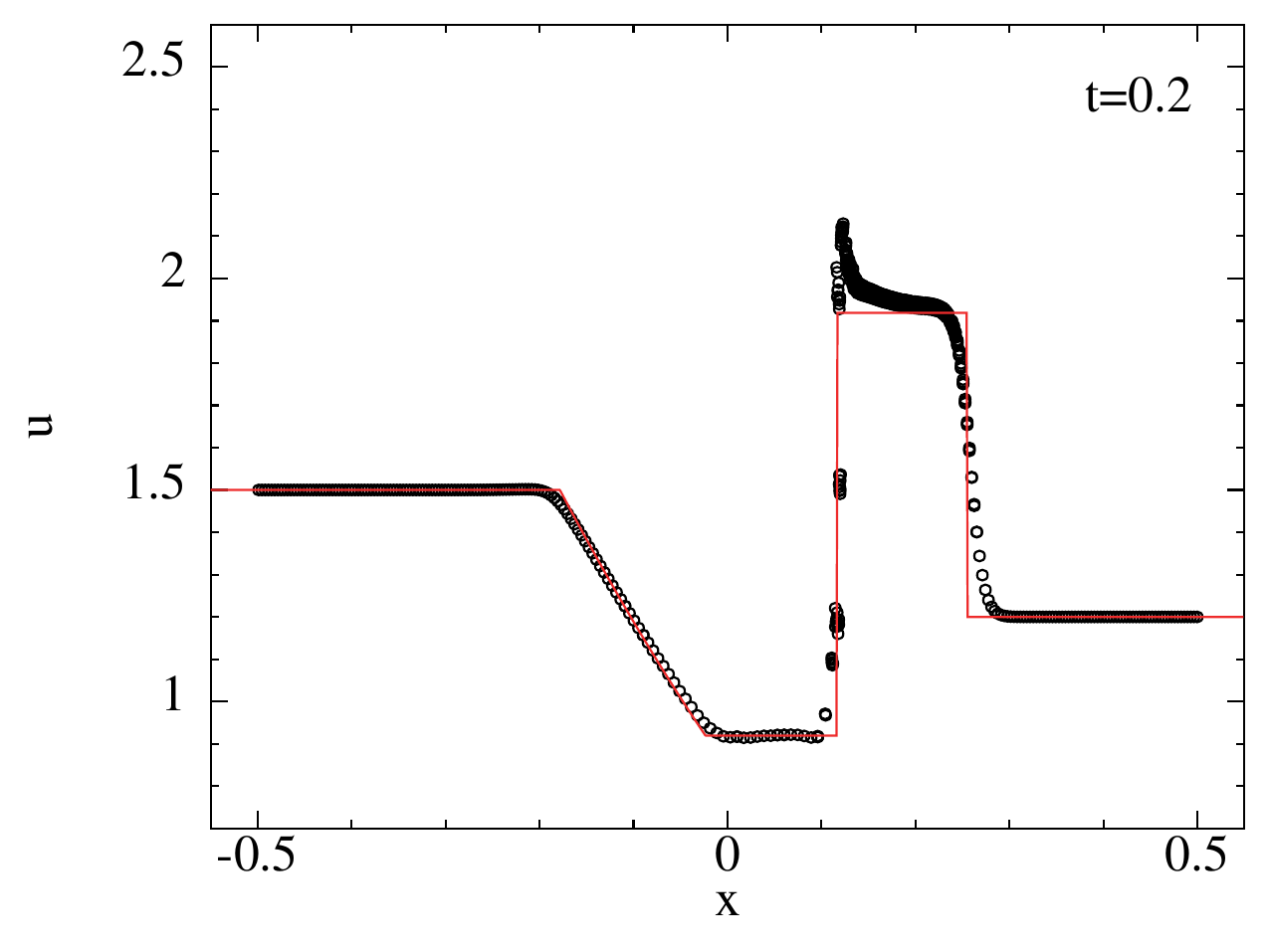}\hfill
    \includegraphics[width=0.47\linewidth]{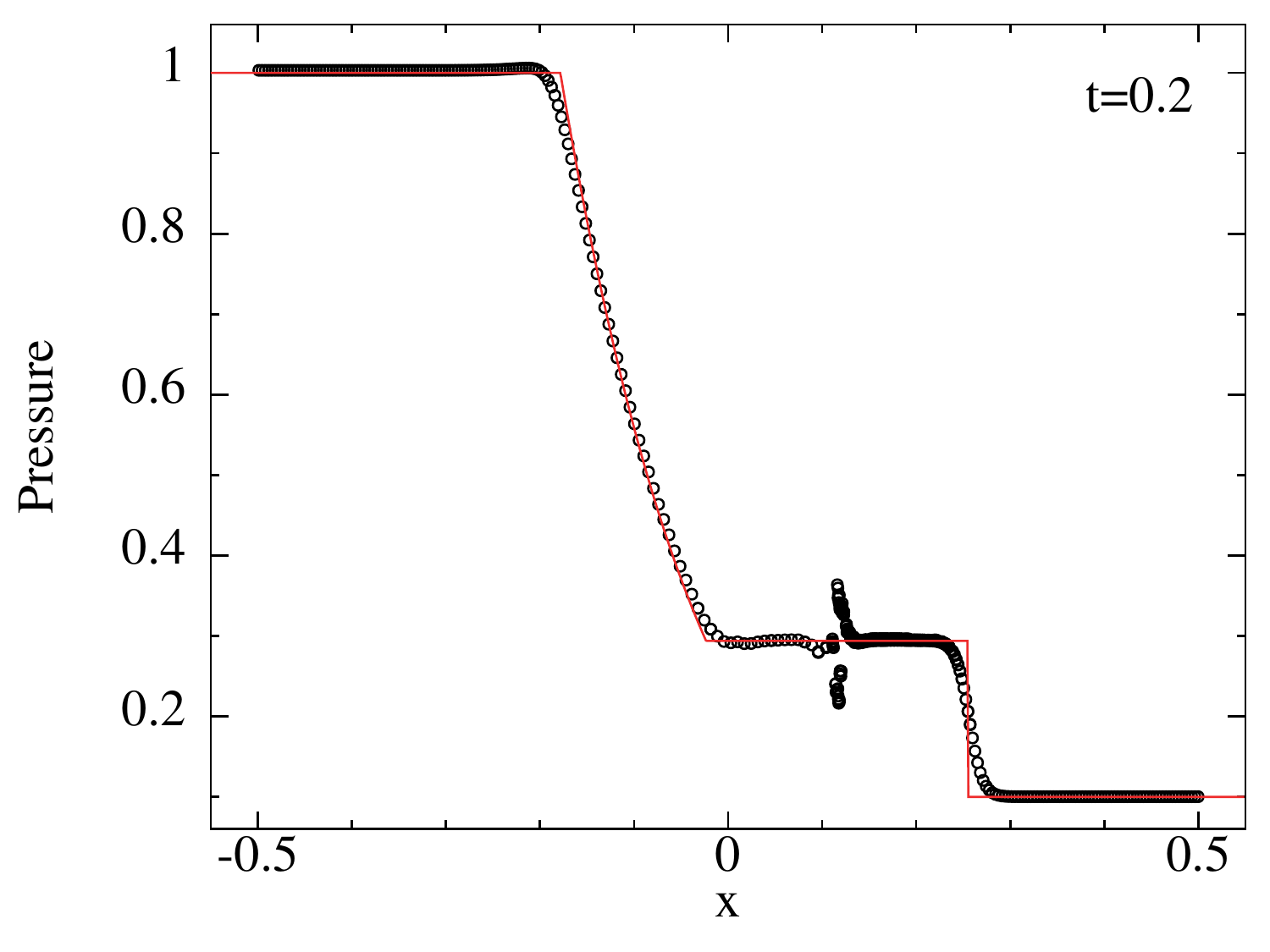}\hfill
    \includegraphics[width=0.47\linewidth]{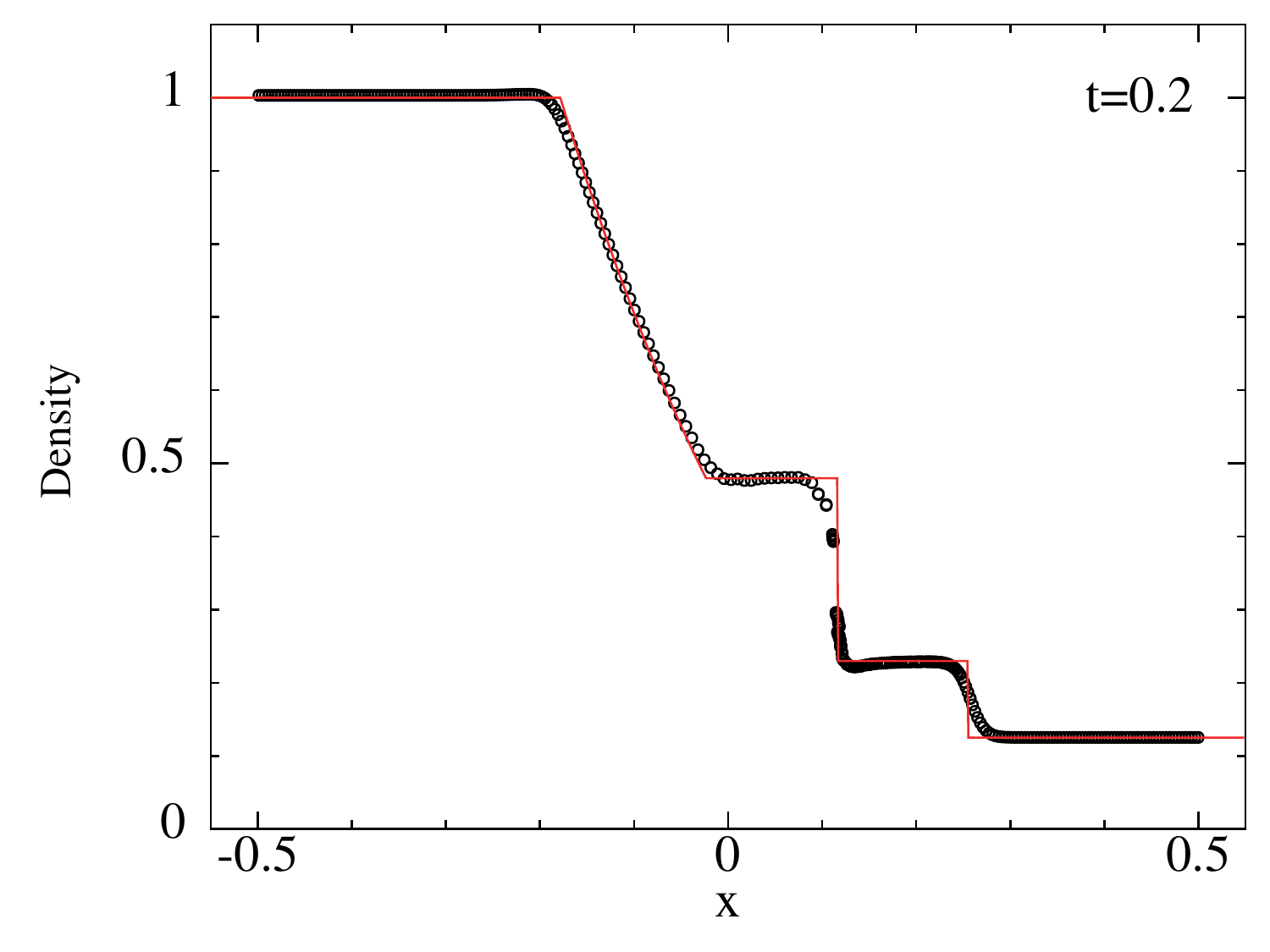}
    \caption{Results from the \textsc{dustyshock} problem with a high drag coefficient and a high dust-to-gas ratio (1:1). The analytic solution is plotted at time $t=0.2$ with a solid red line, the numerical solution is plotted with open circles. From upper left going clockwise plotted are: velocity in $x$ direction, internal energy, gas pressure, and total density. }
    \label{fig:dustyshock}
\end{figure*}

We perform the \textsc{dustyshock} \citep{laibe_dusty_2012a} test here with our implicit scheme at high drag. We use the standard \citet{1978JCoPh..27....1S} shock tube initial conditions. In the gas for $x \leq 0$ $(\rho_\mathrm{g}, v_\mathrm{g}, P) = (1,0,1)$, and for $x>0$ $(\rho_\mathrm{g}, v_\mathrm{g}, P) = (0.125,0,0.125)$. We use an initial dust-to-gas ratio of unity (i.e., $\epsilon = 0.5$), and an adiabatic equation of state where $\gamma = 5/3$.  The domain in 3D is $x,\ y,\ z \in [-0.5,0.5],\ [-0.012,0.012],\ [-0.017,0.017]$ containing 3509 particles, a particle spacing of $ \Delta p \approx 6.84 \times 10^{-3}$. The numerical solution is shown in Fig. \ref{fig:dustyshock} along with the analytic solution for the Sod shock tube in this configuration. The numeric and analytic solutions are in good agreement, with a slight discontinuity in the SPH solution at the contact discontinuity. 

\section{Dusty protostellar collapse}
\label{sec:protostellar-collapse}

As the first application of this new algorithm for solving the dust diffusion equations in dust-as-mixture SPH, we model the dynamics of small dust grains during the protostellar collapse of a molecular cloud core.  We compare the results to those obtained using the semi-implicit dust-as-particles SPH method \citep{loren-aguilar_two-fluid_2014,loren-aguilar_two-fluid_2015} to model the dynamics of the dust grains. 
The dust-as-particles calculations are almost identical to those presented by \citet{bate_dynamics_2017}, except that the initial conditions are modified to set the dust particles to their terminal velocity, since by construction the dust-as-mixture SPH method discussed in this paper assumes the terminal velocity approximation \citep[][took the dust particles to be initially at rest relative to the gas]{bate_dynamics_2017}. We not that this implementation of dust-as-particles does not use the reconstruction of gas and dust velocities of \citet{price_solution_2020}, so as a result linear waves present in the calculation will tend to be overdamped when the dust particle separation becomes too large to resolve the drag length-scale.

In addition to comparing the results from the dust-as-mixture and dust-as-particles calculations, we perform the dust-as-mixture calculations using both explicit and implicit methods for solving the dust diffusion equation to compare their computational performance. 

\subsection{Initial conditions}

Apart from the initial dust velocity being set to its terminal velocity, we replicate the calculations presented in \citet{bate_dynamics_2017}. We set up unstable Bonnor-Ebert spheres of mass $5~\mathrm{M}_\odot$, radius $0.1$ pc, and an inner-to-outer density ratio of 20. We use spherical, reflective boundary conditions. We use initially uniformly rotating clouds, with a rotation rate of $1.012 \times 10^{-13}~\mathrm{rad~s}^{-1}$, and a non-rotating cloud. The magnitude of the ratio of rotational to gravitational potential energy for the rotating cloud is $\beta = 0.08$. 

We use the same equation of state as \citet{bate_dynamics_2017} to model the gas during different phases of collapse, as we do not use radiative transfer. We use a barotropic equation of state for the gas pressure given by
\begin{equation}
P = \left\{ \begin{array}{l l} c_\text{s,0}^2\rho; 		    &  \rho < \rho_\text{c}, \\
c_\text{s,0}^2\rho_\text{c}\left(\rho             /\rho_\text{c}\right)^{7/5};      &  \rho_\text{c} \leq \rho,\end{array}\right.
\end{equation}
where $c_\text{s,0} = 1.87\times 10^4$~cm~s$^{-1}$ is the initial isothermal sound speed of gas with a temperature of 10~K (the mean molecular weight is $\mu=2.38$), and $\rho_\text{c} = 10^{-13}$ g~cm$^{-3}$. As in  \citeauthor{bate_dynamics_2017}, our calculations only follow the collapse until shortly after the density exceeds $\rho_{\rm c}$, so the gas almost always isothermal at 10~K, except in the centre at the end of the calculations when the temperature begins to rise.

In the dust-as-particles calculations we use 1 million SPH particles to model the gas, and 300,000 to model the dust.  Comparatively in the dust-as-mixture calculations we use 1 million SPH particles that model both gas and dust as one fluid in a dust-as-mixture model. A uniform initial dust-to-gas ratio of 1:100 is used for all models. Using identical numbers of SPH particles to model the gas components in both sets of calculations means the gas resolutions are the same and the only difference between the calculations is due to the dust implementation.

\subsection{Results}

\begin{figure*}
    \centering
    \includegraphics[width=0.45\linewidth]{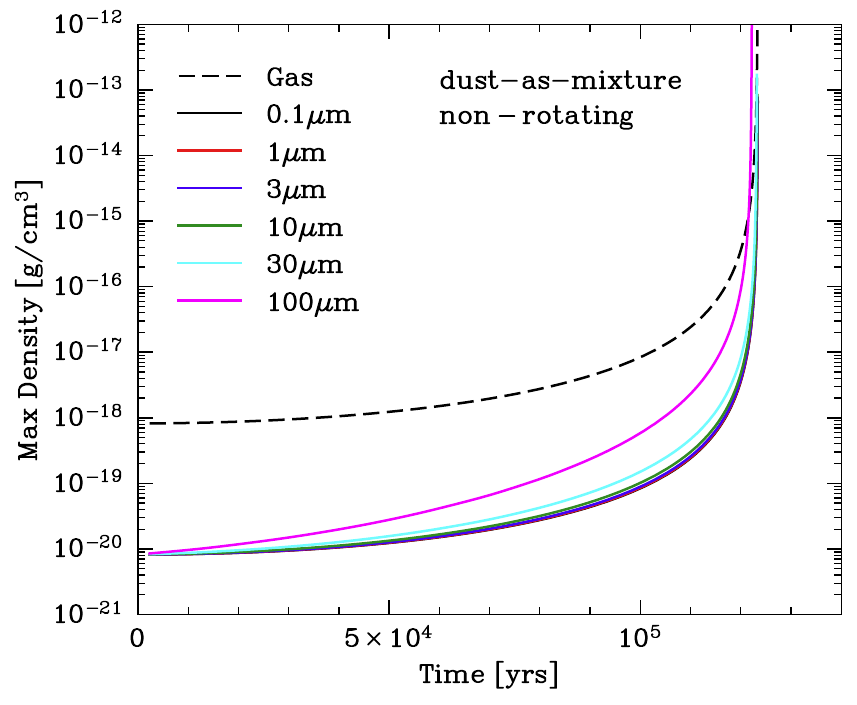}
    \includegraphics[width=0.45\linewidth]{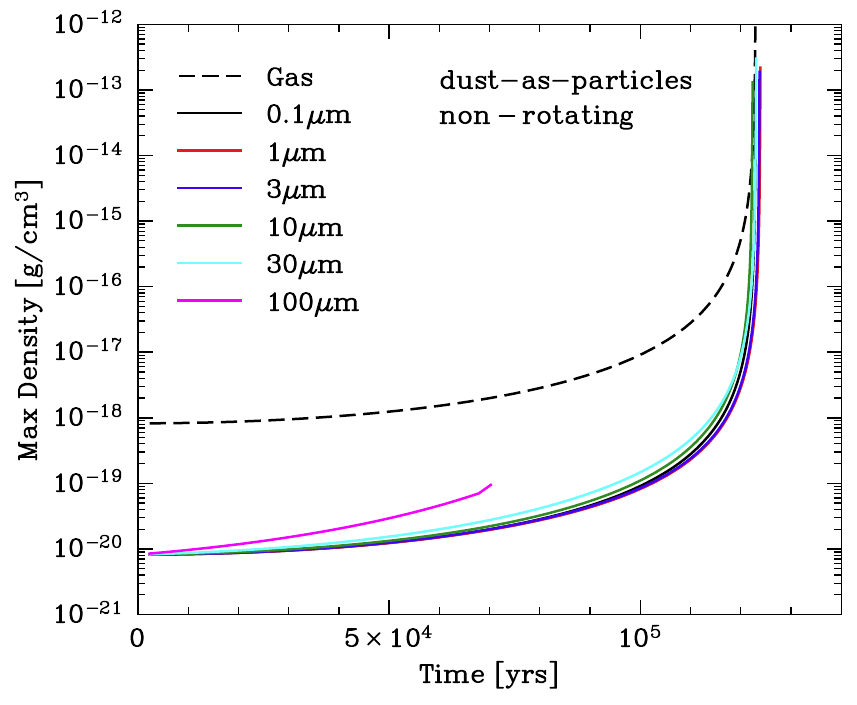}
    \caption{Time evolution of the central density of gas and dust during the collapse of non-rotating molecular cloud cores performed using  in dust-as-mixture (left) and dust-as-particles (right) methods. There is a monatonic increase in the maximum density of dust with time. The dust-as-mixture and dust-as-particles solutions are similar, except for the largest dust grains. In the dust-as-particles $100 \mu \mathrm{m}$ calculation, the dust in the outer parts of the cloud is collapses more rapidly than in the inner regions, resulting in a `pile up' of dust.  For this case, the terminal velocity approximation of the dust-as-mixture method is only comparable until $t\approx 70,000$~yrs, so we only plot the dust-as-particles solution to this point. The gas density is from the $0.1 \mu \mathrm{m}$ calculations.}
    \label{fig:non-rotating-centdens}
\end{figure*}

\begin{figure*}
    \centering
    \includegraphics[width=0.45\linewidth]{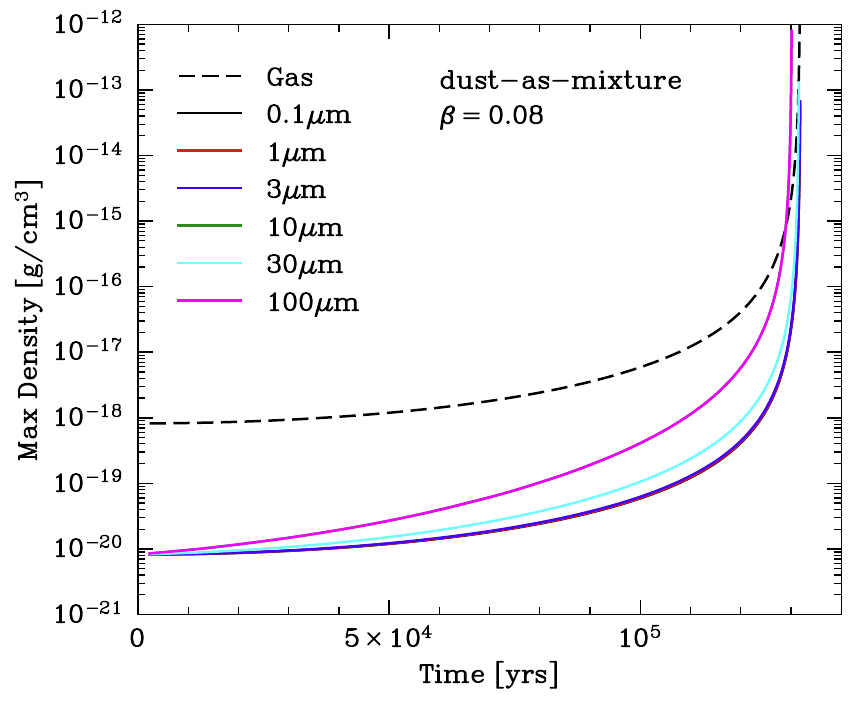}
    \includegraphics[width=0.45\linewidth]{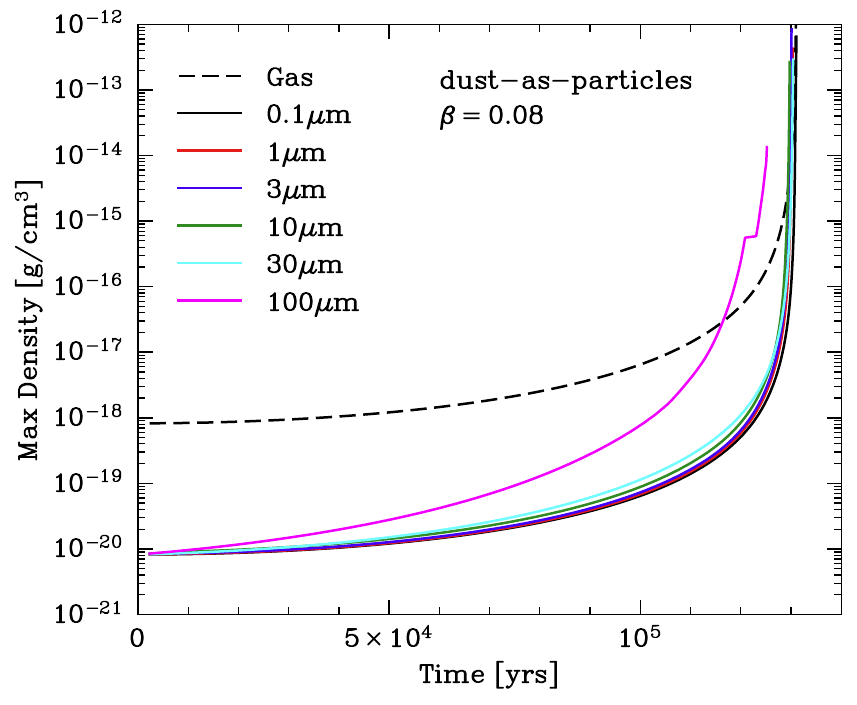}
    \caption{Time evolution of the central density of gas and dust in both rotating calculations in dust-as-mixture (left) and dust-as-particles (right). Due to the rotation of the cloud the formation of the hydrostatic core is delayed. The main differences between the two methods is with the largest grains; it is with this size of grain where the underlying assumptions of dust-as-mixture begin to breakdown and a dust-as-particles solution is desirable. The gas density is from the $0.1 \mu \mathrm{m}$ calculations.}
    \label{fig:centraldensrot}
\end{figure*}

\begin{figure*}
    \centering
    \includegraphics[width=0.9\linewidth]{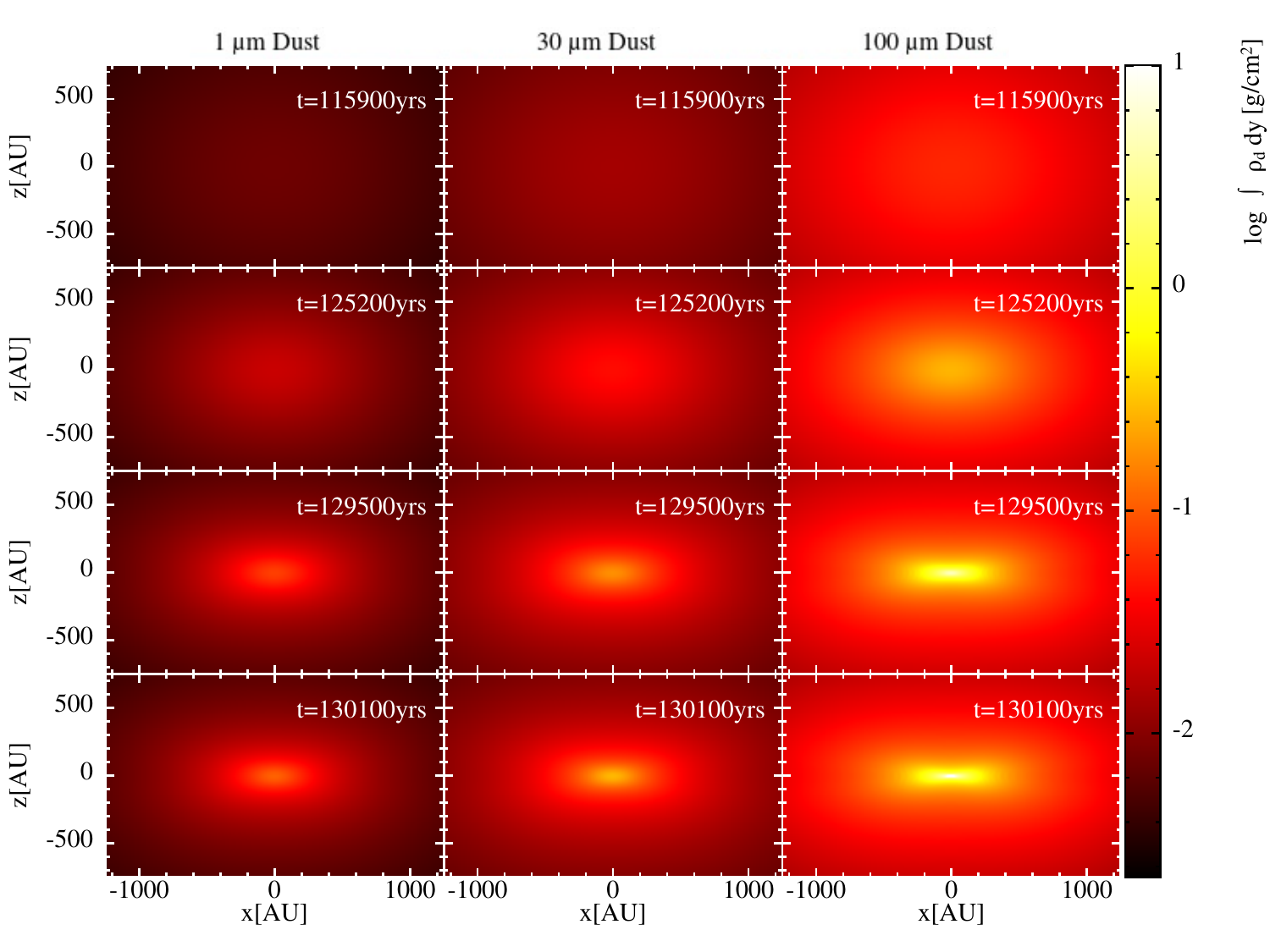}
    \caption{Evolution of dust grain column density viewed in the plane of rotation for  calculations of molecular cloud core collapse using the dust-as-mixture method. Plotted from left to right are calculations with dust grain sizes $r_\mathrm{s} = 1\mu \mathrm{m},\ 30\mu \mathrm{m},\ 100\mu \mathrm{m}$. The colour bar is set such that it is the same in each panel of the plot. The small dust grains are coupled to the gas and follow the gas well. The larger grains exhibit different behaviour whereby they collapse more quickly resulting in the central oblate region having a higher dust density. The central obliquity is due to the overall rotation of the gas cloud. }
    \label{fig:dust_rho_B008}
\end{figure*}

\begin{figure*}
    \centering
    \includegraphics[width=0.9\linewidth]{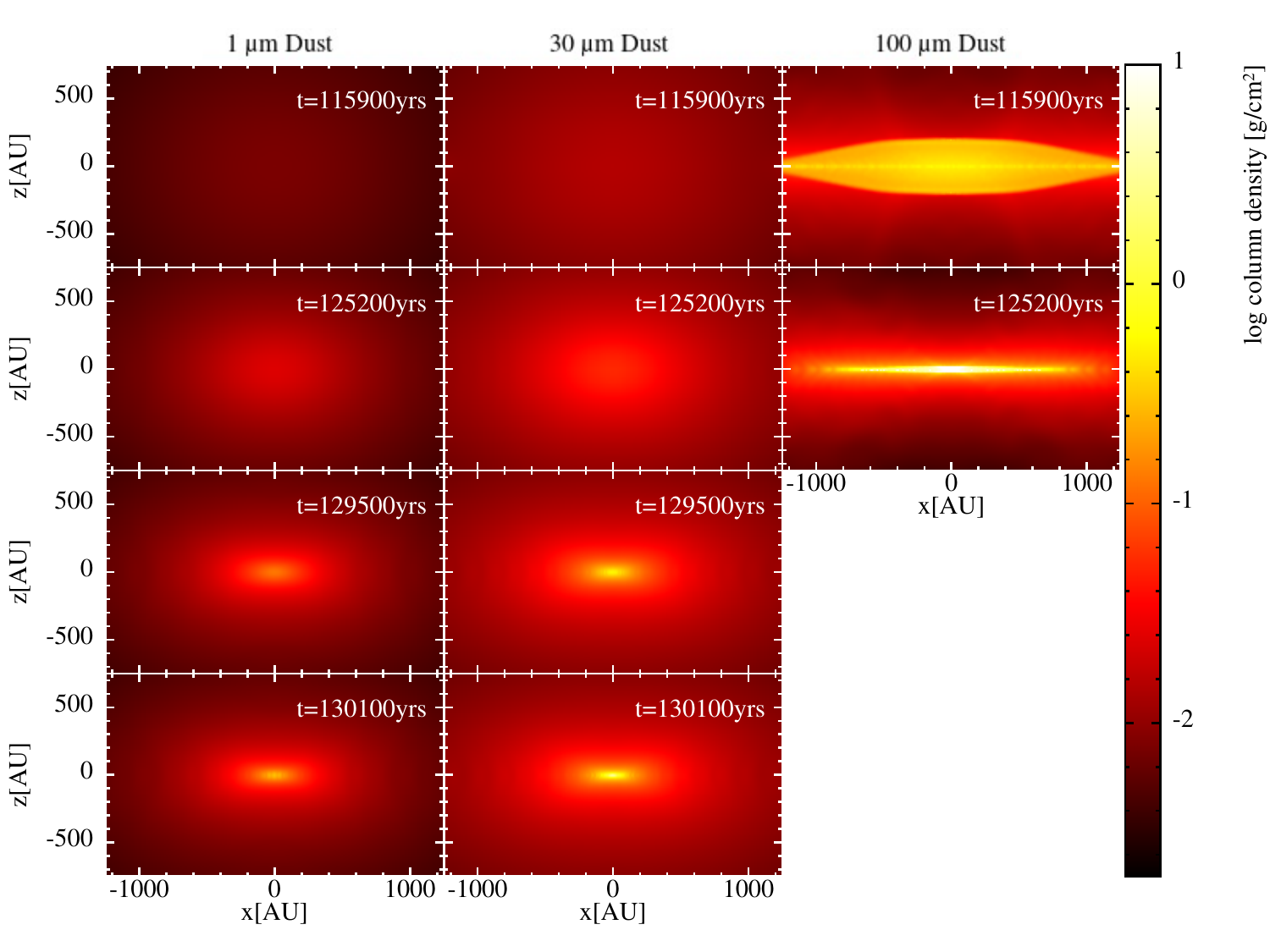}
    \caption{Evolution of dust grain column density in the plane of cloud rotation for calculations of molecular cloud core collapse using the dust-as-particles method. Shown from left to right are calculations with dust grain sizes $r_\mathrm{s} = 1\mu \mathrm{m},\ 30\mu \mathrm{m},\ 100\mu \mathrm{m}$. The colour bar is set such that it is the same in each panel of the plot. In the $100 \mu \mathrm{m}$ calculation the dust grains from the outer regions of the cloud collapse faster than the rest of the grains causing a pile up of grains that pass through each other in the centre of the cloud.  These subsequently settle into a dense flat disc, whereupon the calculation grinds to a halt.}
    \label{fig:dust_2F_rho_B008}
\end{figure*}

Here we present the results of the protostellar collapse of a non-rotating cloud and a rotating cloud with $\beta = 0.08$ using both the dust-as-mixture and -particles SPH methods.

In Fig. \ref{fig:non-rotating-centdens} we show the time evolution of the densities of gas and dust at the centre of the collapsing non-rotating clouds. In the dust-as-mixture calculations (left panel) there is a monotonic increase in the density of dust, and the rate of increase is greater for larger dust grain sizes due to the weaker coupling to the gas, although there is little difference for grains of size $r_\mathrm{s} \lesssim 10 \mu \text{m}$. The results are very similar in the equivalent dust-as-particles calculations (right panels), except for the largest grain size. For very small dust grain sizes there is good agreement between the dust-as-mixture and dust-as-particles results that only starts to break down very late in the collapse at very high densities as the first hydrostatic core \citep{larson_1969MNRAS.145..271L} forms; in the the dust-as-particles calculations (unphysical) dust particle clumping occurs late in the collapse.  For intermediate dust grain sizes ($r_\mathrm{s} \approx 10-30~\mu$m) the central dust density tends to be slightly higher for the dust-as-particles calculations than the dust-as-mixture calculations in the latter third of the collapse, again due to slight dust clumping when using the dust-as-particles method.

In the dust-as-particles calculation with dust grains of size $r_\mathrm{s} = 100 \mu \mathrm{m}$, the initial terminal infall velocities of the dust in the outer regions of the cloud are much higher than in the centre. Therefore, the dust in the outer regions collapses much faster than the dust in dense inner regions causing a `pile-up' as the dust collapses.  This is an extreme case for which the terminal velocity approximation assumed by the dust-as-mixture model quickly breaks down. Therefore, for this case, we present the dust-as-particles results in Fig. \ref{fig:non-rotating-centdens} only until the dust from the outer regions of the cloud catches up with that from the inner part of the cloud. The results between the dust-as-mixture and dust-as-particles for $100 \mu \mathrm{m}$ grains are similar until this point ($t \approx 70,000$ yrs).

In Fig. \ref{fig:centraldensrot} we show the time evolution of the gas and dust densities at the centre of the rotating clouds. In the calculations with dust grains $r_\mathrm{s} <100 \mu \mathrm{m}$ there is little difference in the central densities of dust and gas between the two methods. Again, the dust-as-particles calculations are prone to some dust clumping.  Also again, the largest grains ($100\mu\text{m}$) are less well coupled to the gas and so collapse much more quickly than the small grains.  In the rotating case the collapse happens slower than in the non-rotating case, as expected due to centrifugal forces. But again, the dynamics of the large dust grains is better captured using the dust-as-particles method as in this case the terminal velocity approximation is not appropriate.

In Fig. \ref{fig:dust_rho_B008} we show snapshots of the evolution of the column density of dust in the rotating protostellar collapse calculations containing dust with grain sizes of $r_\mathrm{s} = 1\mu \mathrm{m},\ 30 \mu \mathrm{m},\ 100 \mu \mathrm{m}$. The region shown is the centre of the cloud in the plane of rotation. The small dust grains are tightly coupled to the gas (the morphology of the gas column density looks almost identical to the 1$\mu$m dust morphology), whereas the larger grains are not as well coupled and collapse more quickly. This leads to an enhancement in dust-to-gas ratio when larger dust grains are present. The $100 \mu \mathrm{m}$ dust grains collapse fast enough to form a disc-like structure well before the stellar core forms, a consequence of the rotation rate of the cloud. A qualitatively similar result was found by \citet{bate_dynamics_2017}, shown in  Fig. 3 of their work. We show the equivalent plot of Fig. \ref{fig:dust_rho_B008} for the dust-as-particles calculations in Fig. \ref{fig:dust_2F_rho_B008}. The dust density for grain sizes 1 and 30$\mu \mathrm{m}$ are similar in both the one- and dust-as-particles calculations, with the exception that in the dust-as-particles calculation the density at the centre of the cloud is slightly denser than in the dust-as-mixture at very late times. The 100$\mu \mathrm{m}$ dust grain calculation using dust-as-particles looks vastly different. In this calculation the dust grains in the outer parts of the cloud are very decoupled from the gas and have such a large initial terminal velocities that they are able to collapse to the centre of the cloud very quickly.  Moreover, because the gas drag on them is so weak, even when they get to the denser inner regions the drag is insufficient to slow them to their local terminal velocity so they continue to collapse more quickly than in the dust-as-mixture calculation.  This is a case that can not be modelled accurately using the dust-as-mixture terminal velocity approximation. We note that this is an extreme case -- such large dust grains are not thought to be abundant in molecular cloud cores.  Interstellar dust grains are typically thought to be sub-micron in size.

\subsection{Comparison with explicit timestepping}

In this section we compare the computational performance of the explicit timestepping algorithm of the dust evolution equation as described in \citet{ballabio_dust_fraction_2018} with the implicit algorithm described in this work. In their formulation \citet{ballabio_dust_fraction_2018} impose a criterion upon the stopping time of the dust to prevent numerical artefacts in regions of low dust-to-gas ratio, namely 
\begin{equation}
    \Tilde{t}_\mathrm{s} = \min(t_\mathrm{s},h/c_\mathrm{s}).
\end{equation}
The reasoning for adopting this stopping time is to prevent infeasibly slow calculations by limiting the mass flux of particles in regions where there are steep dust density gradients, i.e. at the edge of discs. However this artificially alters the drag and thus the effective size of the dust grains and should be used with caution.

As an example of the problems that altering the stopping time can give, we repeat some of the dust-as-mixture non-rotating protostellar collapse calculations from Section \ref{sec:protostellar-collapse} using explicit time stepping with the stopping time limiter turned on and off.  We compare the results and performance with our implicit algorithm without a stopping time limiter.

\begin{figure}
    \centering
    \includegraphics[width=\linewidth]{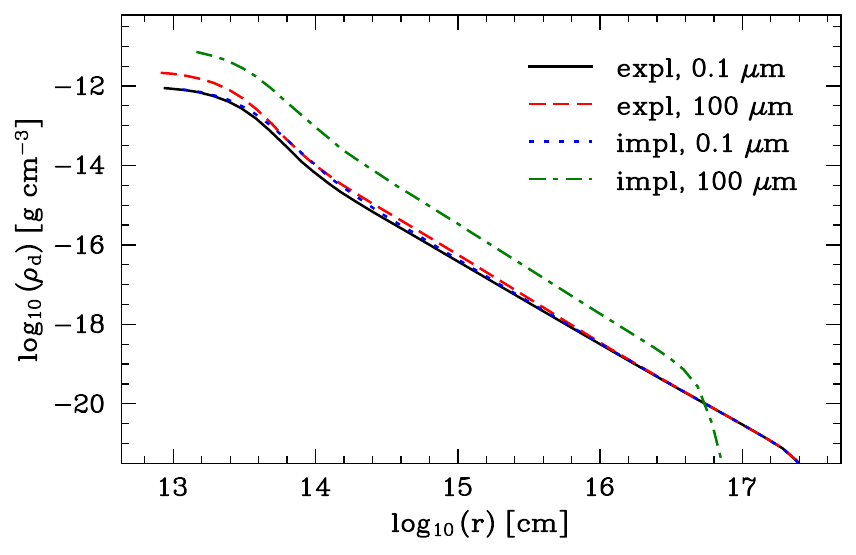}
    \caption{Dust density as a function of radius for dust grain sizes $0.1$ and $100 \mu \mathrm{m}$ in non-rotating  calculations of protostellar collapse when the peak gas density reaches of $10^{-10} \text{g cm}^{-3}$, just after the first hydrostatic core begins to form. The `expl' lines denote the dust density from explicit dust-as-mixture calculations with the stopping time limited applied, and `impl' lines denote the equivalent calculations from implicit dust-as-mixture calculations (without a stopping time limiter). Due to the action of the stopping time limiter the dust density profiles for the two explicit calculations with different grain sizes are very similar, whereas with the implicit method the large grains have evacuated the outer parts of the cloud and migrated much more strongly inward.}
    \label{fig:tl_1fluid_collapse}
\end{figure}

In Fig. \ref{fig:tl_1fluid_collapse} we show the distribution of dust as a function of radius in the collapsing cloud when the maximum gas density reaches $\rho = 10^{-10}\ \mathrm{g~cm}^{-3}$ using both the explicit and implicit methods.  The explicit results have the stopping time limiter applied, but no stopping time limiter is used for the implicit calculations.  We give results for dust grain sizes $r_\mathrm{s} = 0.1,\ 100 \mu \mathrm{m}$. In the explicit, stopping time limited calculations the dust density distribution for both grain sizes is very similar throughout the cloud, and identical in the outer regions where the gas density is low and the stopping time is the longest (i.e. where the stopping time limiter is most likely to be active). The implicit result for the $0.1\mu$m grain size is very similar to the explicit result.  However, in the implicit calculation with large grains the dust density is very different. In the outer parts of the cloud, the dust has fallen inwards leaving no dust in the outer parts of the cloud and producing much higher dust densities in the inner regions. Near the centre of the clouds the dust density distributions for the two explicit calculations with the stopping time limiter differ slightly, with the larger grains falling faster leading to a somewhat higher density than the small grains, due to the stopping time limited turning off at the higher gas densities. However the density of the larger grains here is still substantially underestimated compared with the implicit calculation. The large grains in the explicit calculation behave as if they were much smaller due to the timestep limiter. This is why extreme caution is needed if a stopping time limiter is used -- a plausible result is obtained for the $r_\mathrm{s} = 100 \mu \mathrm{m}$ explicit calculation with the stopping time limiter, but it is a physically incorrect result. 


To compare the computational performance between the explicit algorithm and the implicit algorithm, we also ran an explicit calculation without any stopping time limiter for the $r_\mathrm{s} = 0.1 \mu \mathrm{m}$ dust grains (a typical size for interstellar dust grains). This increased the calculation time so significantly that it was unrealistic to allow the gas density to reach $\rho = 10 ^ {-10} \mathrm{g~cm} ^ {-3}$. We estimate that the implicit algorithm is at least two orders of magnitude (i.e., several hundred times) faster than the explicit algorithm at calculating this protostellar collapse case, because the implicit calculation takes standard hydrodynamical timesteps and does not require the use of a dust-as-mixture dust timestep.  When comparing the dust density distributions of these two calculations they are almost identical; our algorithm yields the same result as the explicit algorithm without limiting the stopping time, but does so much more quickly.  When using the explicit method with the stopping time limiter for the $0.1\mu$m grain size, the explicit calculation was only about 5 percent faster than the implicit method. 

\section{Protoplanetary discs}\label{sec:ppd}

In this section we present the results from protoplanetary disc simulations similar to those of \citet{ballabio_dust_fraction_2018}. The calculations include an embedded planet to study the diffusivity effects of gradients in dust fraction introduced by planet-disc interactions. The idea is that as the planet orbits, gas and dust are flung out into the outer regions of the disc where there are steep dust fraction gradients. It is in these environments that \citeauthor{ballabio_dust_fraction_2018} found that numerical artifacts in the dust can arise. This is most notable when using earlier formulations that evolved the dust fraction or the dust-to-gas ratio.

For our planet-in-disc test calculations, the initial radial extend of the gas is $r \in [25,200]$ au and the dust initially extends to 90\% of the gas radius.  A locally-isothermal equation of state is used, with a radial temperature proportional to $r^{-1/2}$ so that ratio of the disc scale height to radius is a constant, $H/r = 0.1$. The central object is modelled using a gravitational potential of mass $1~\mathrm{M}_\odot$. The surface density of the disc has an initial profile proportional to $r^{-1/2}$.  The mass of the disc is $0.0348~\mathrm{M}_\odot$ and the dust-to-gas ratio is set at 0.01. We used 1 million SPH particles to model the disc. The disc contains a 6 Jupiter-mass (M$_\mathrm{J}$) planet at a radius of $140$ au. The planet is modelled using a sink particle \citep
{bate_sphng_1995MNRAS.277..362B}, with an accretion radius of 10 au. The sink particle only interacts with the gas via gravity and accretion, and the trajectory of the sink particle is integrated with the same Runge-Kutta-Fehlberg integrator that is used for the SPH particles, but with a much lower tolerance.

\begin{figure*}
    \centering
    \includegraphics[width=\linewidth]{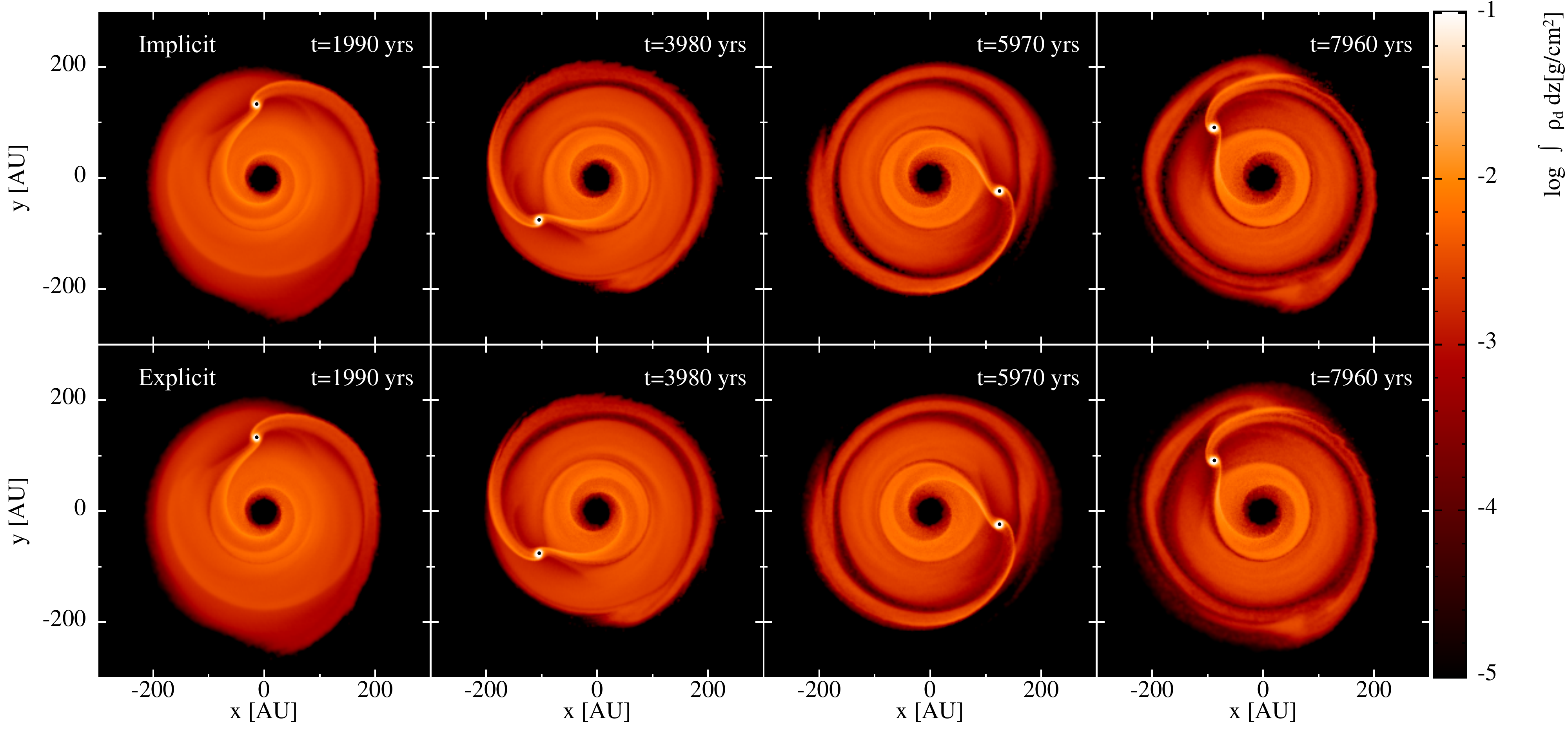}
    \caption{The time evolution of the column density of dust in a protoplanetary disc containing a 6 M$_\mathrm{J}$ planet. \textit{Upper panels:} Results using the implicit dust-as-mixture dust algorithm. \textit{Lower panels:} Results using the explicit dust-as-mixture dust algorithm with the stopping time limiter. Both calculations contain $100 \mu \mathrm{m}$ dust grains and use the same initial conditions.  Similar results are obtained, but the implicit method gives slightly `sharper' edges to the dust in the gap produced by the planet and at the outer edge of the disc.}
    \label{fig:dust_comp}
\end{figure*}

In Fig. \ref{fig:dust_comp} we show the time evolution of the dust column density within the protoplanetary disc containing dust grains of size $100 \mu \mathrm{m}$. The upper panels show the results from a calculation using the implicit dust-as-mixture algorithm to evolve the dust, the low panels show an identical calculation but using the explicit dust-as-mixture algorithm with the stopping time limiter. 

The overall dust density distributions are almost identical. However, there are some differences in regions of high dust density gradient at the dust gap created by the planet and at the outer edge of the dust disc. In the explicit dust-as-mixture calculation the dust density is more smoothed out, whereas in the implicit dust-as-mixture calculation the edges of the dust disc are more sharply defined. This is most easily spotted in the gap in the dust; the dust does not diffuse into the gap as much in the implicit calculation.

\begin{figure}
    \centering
    \includegraphics[width=\linewidth]{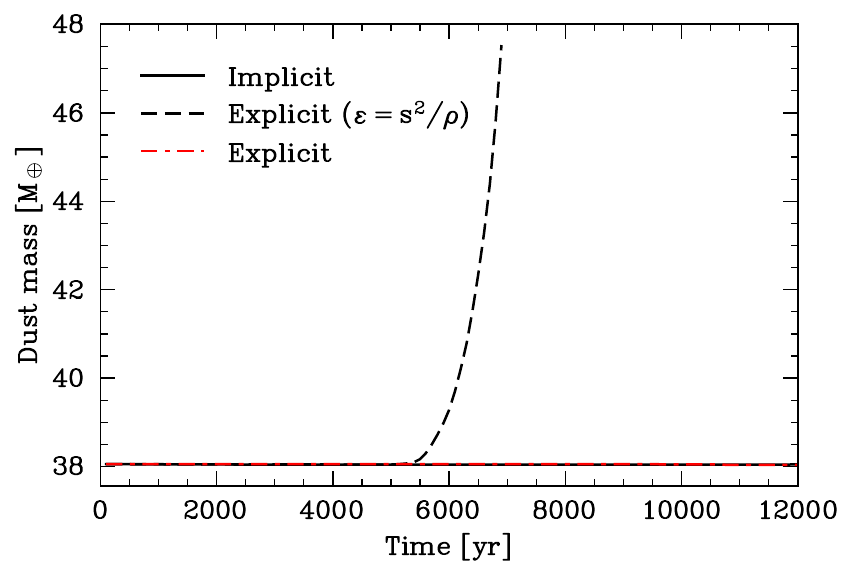}
    \includegraphics[width=\linewidth]{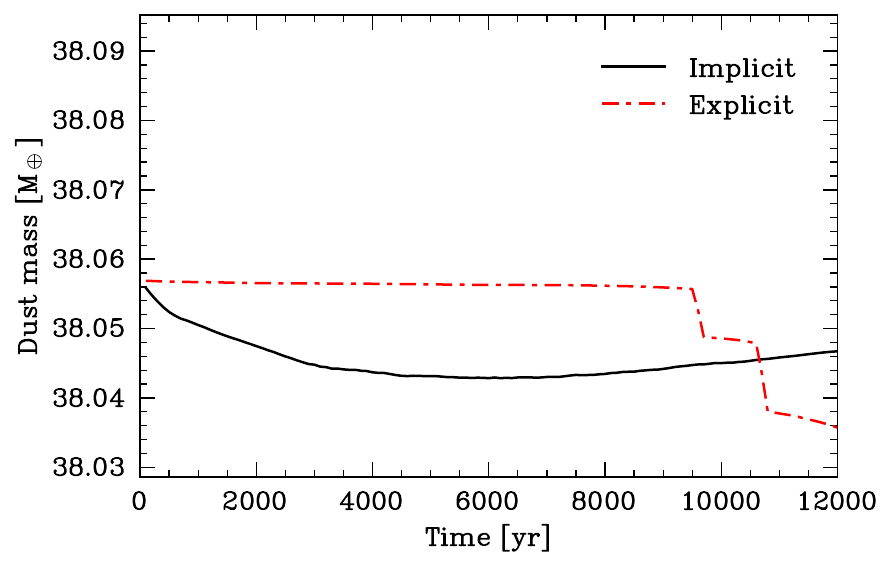}
    \caption{Dust mass evolution in a disc with an embedded planet. \textit{Upper}: Dust mass evolution for the implicit and explicit calculations using the \citet{ballabio_dust_fraction_2018} dust variable, and an explicit calculation using an alternate dust variable. \textit{Lower}: A zoom in of the results for our implicit dust evolution algorithm and the explicit dust evolution algorithm, both using the \citeauthor{ballabio_dust_fraction_2018} parameterisation. Both of these calculations conserve dust mass to better than 0.1 percent.}
    \label{fig:dustconservation}
\end{figure}

We tested the mass conservation of the implicit dust-as-mixture algorithm and find it performs similarly to the explicit stopping time limited algorithm at conserving mass. We performed a similar calculation to the one described above except that we used only 200,000 SPH particles and we modelled the planet and disc in the same way as \citet{ayliffe_gas_2009} where gas and dust falls on to and is bound to a planetary surface and remains in the calculation (rather than being accreted by a sink particle).  Since all of the mass remains in the calculation (as opposed to being accreted by the sink particle) this allows us to measure more precisely the mass conservation of the respective dust schemes. In the upper panel of Fig. \ref{fig:dustconservation} we show that we recover the result reported by  \citet{ballabio_dust_fraction_2018} in which the parameterisation of dust fraction given by \citet{price_small_grains_2015MNRAS.451..813P}, $\epsilon = s^2 / \rho$, does not conserve mass (the dust mass increases with time). In contrast, our implicit dust algorithm maintains the dust mass as well as the explicit algorithm with the \citet{ballabio_dust_fraction_2018} dust parameterisation and stopping time limiter. In the lower panel of Fig. \ref{fig:dustconservation} we focus on just the implicit and explicit calculations to compare how well they converse mass against each other. Our implicit calculation slowly loses a small fraction of mass ($\approx$ 0.02 \%) initially and then approximately maintains this mass throughout the remainder of the calculation. In the case of the explicit calculation, the initial dust mass very slightly decreases through most of the calculation, but then loses a small fraction of mass towards the end ($\approx$ 0.06\%). Overall we find that the dust mass conservation of the two different algorithms is similarly good.

\section{Negativity of the dust parameter}

Whilst the parameterisation of the dust fraction as given by in equation \ref{eq:Ballabio} ensures that it remains positive and less than one, there is no such constraint on the dust evolution variable, $s$, itself. During calculations, as the parameter is evolved, it can become negative. The dust fraction itself remains limited to between zero and unity, but a negative dust variable can still produce unwanted, nonphysical effects in the dust fraction. These effects typically manifest as dust seemingly `leaking out' into regions of low dust fraction, at the edges of discs for example. This was pointed out by \citet{price_small_grains_2015MNRAS.451..813P} for earlier versions of the dust parameterisation.  They found that the most effective way to counter this issue was to set the dust fraction of the problem particles to zero. 

In what follows, we show that the problem of the dust variable going negative can still occur with the newer dust parameterisation of \citet{ballabio_dust_fraction_2018}. Here we demonstrate the effects of a negative dust variable in both an idealised test case and in a more physically motivated fully 3D simulation.  We then demonstrate that another advantage of the implicit dust method is that it avoids this problem entirely.

\subsection{Dust settling test}

For the idealised test case, we use the dust settling test of \citet{price_small_grains_2015MNRAS.451..813P} which considers vertical dust settling in a Cartesian box with an acceleration in the `vertical' (z) direction that mimics a small piece of a protoplanetary disc (i.e., the isothermal gas has a Gaussian density distribution vertically centred on $z=0$).  A full description of the set up the test is given by \citet{price_small_grains_2015MNRAS.451..813P} and will not be repeated here.  In our case, we start the test with a dust-to-gas ratio of 1:100 and use a dust grain size of 100$\mu$m.  The orbital period is 353 years and we show the state of the dust layer after 20 orbits.

\begin{figure}
    \centering

    \begin{subfigure}[b]{0.45\textwidth}
        \includegraphics[width=\linewidth]{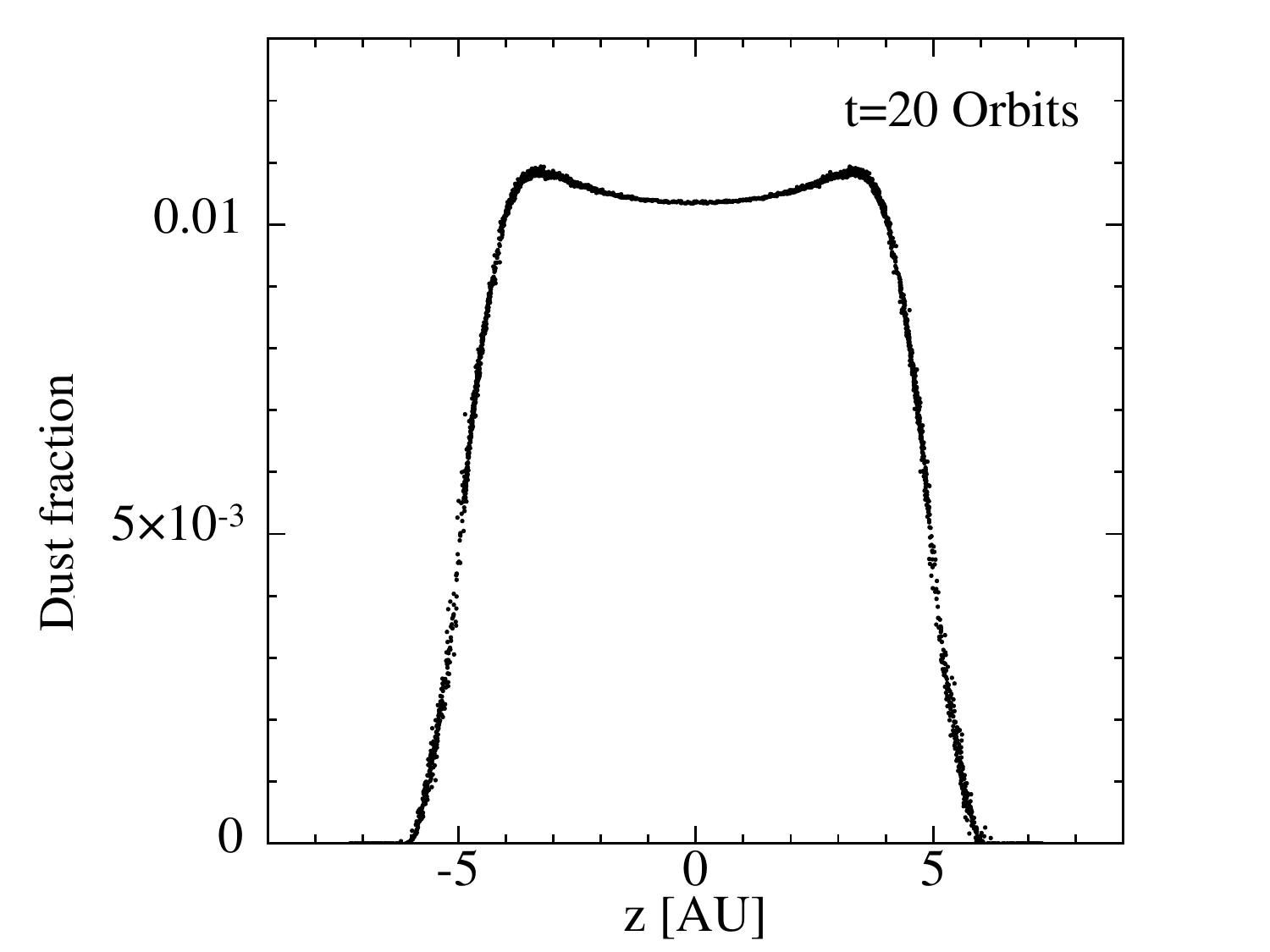}
    \end{subfigure}
    \begin{subfigure}[b]{0.45\textwidth}
        \includegraphics[width=\linewidth]{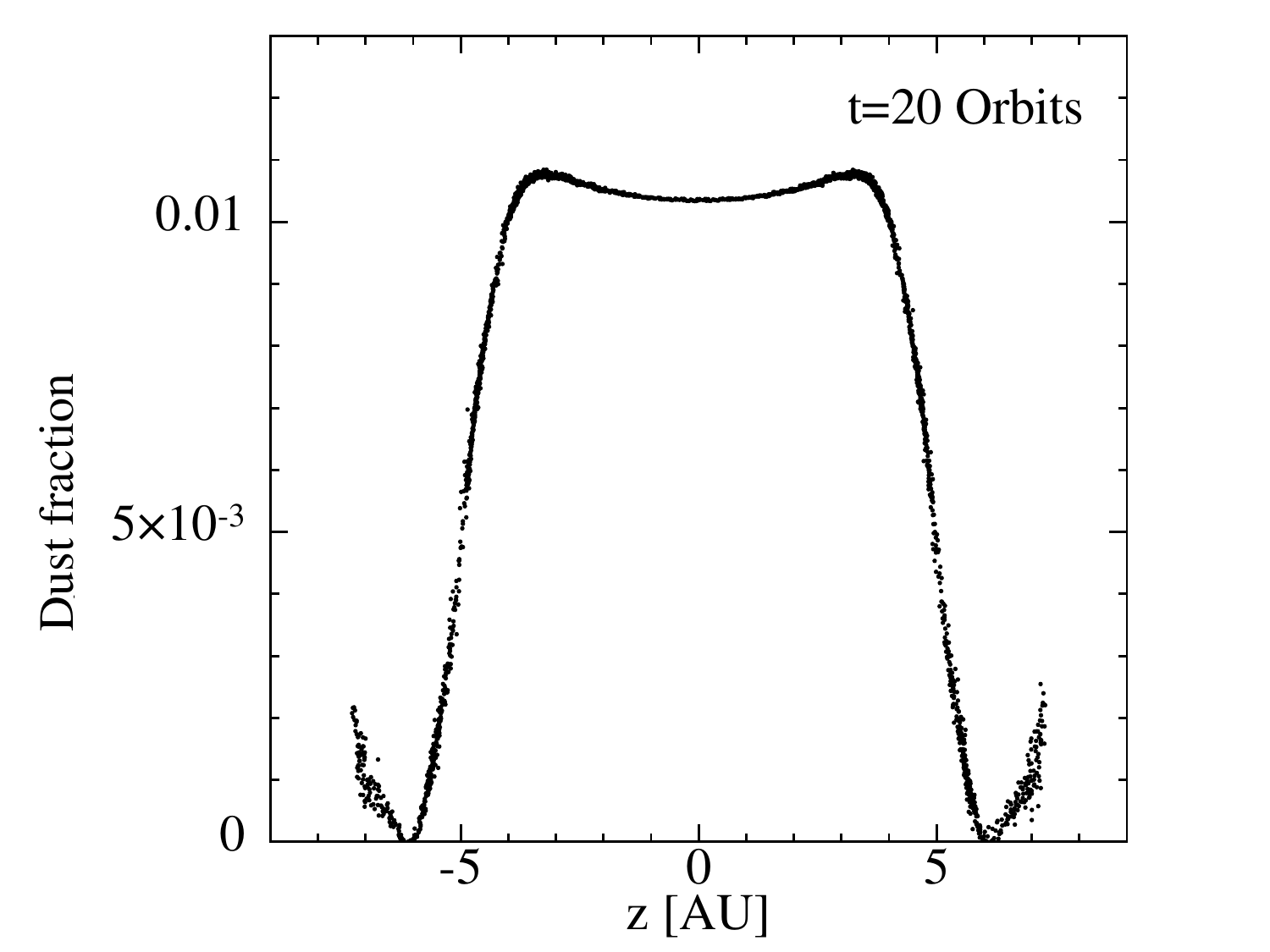}
    \end{subfigure}
    \caption{The dust settling test with 100$\mu$m dust grains after 20 orbital periods. \textit{Upper panel}: dust fraction as a function of vertical  height in the disc using the implicit method to evolve the dust. \textit{Lower panel}: the same except using the explicit method. Using the explicit method, dust `leaks' out of the dust layer into the low-density atmosphere of the disc ($|z|>6$ au in the above panels).  Using the implicit method gives the same dust profile near the midplane of the disc, but outside the dust layer where the dust fraction should be zero, it is zero as it should be.}
    \label{fig:dustsettle}
\end{figure}

As the disc evolves, the dust falls out of the low-density surface regions of the disc to create a dust layer with steep dust gradients at its surface. We show our results in Fig. \ref{fig:dustsettle} using both the explicit and implicit methods. The problem of the dust variable becoming negative occurs at this sharp gradient at the `surface' of the dust layer. This is due to the way exact mass conservation of the dust-as-mixture equations is ensured by pairwise SPH particle interactions.  The equations `take' dust from one SPH particle and `give' it to another.  However, if the dust variable of the donor particle is small this can lead to the dust variable of the donor particle becoming negative. Due to the definition of the dust fraction, a negative dust variable still gives a positive dust fraction. However, this means that in regions where no dust is expected, dust diffuses out into these regions in an unphysical manner (lower panel of Fig. \ref{fig:dustsettle}).  This can be avoided in the explicit method by explicitly forcing the dust variable to remain positive, but this comes at the expense of sacrificing dust mass conservation.

Using our implicit method, in which the solution to the dust diffusion equation is iterated using the Gauss-Seidel method, the dust variable remains positive and the apparent diffusion of the dust out of the dust layer is avoided (upper panel of Fig. \ref{fig:dustsettle}).  

\subsection{Protoplanetary disc}

\begin{figure*}
    \centering
    \includegraphics[width=\linewidth]{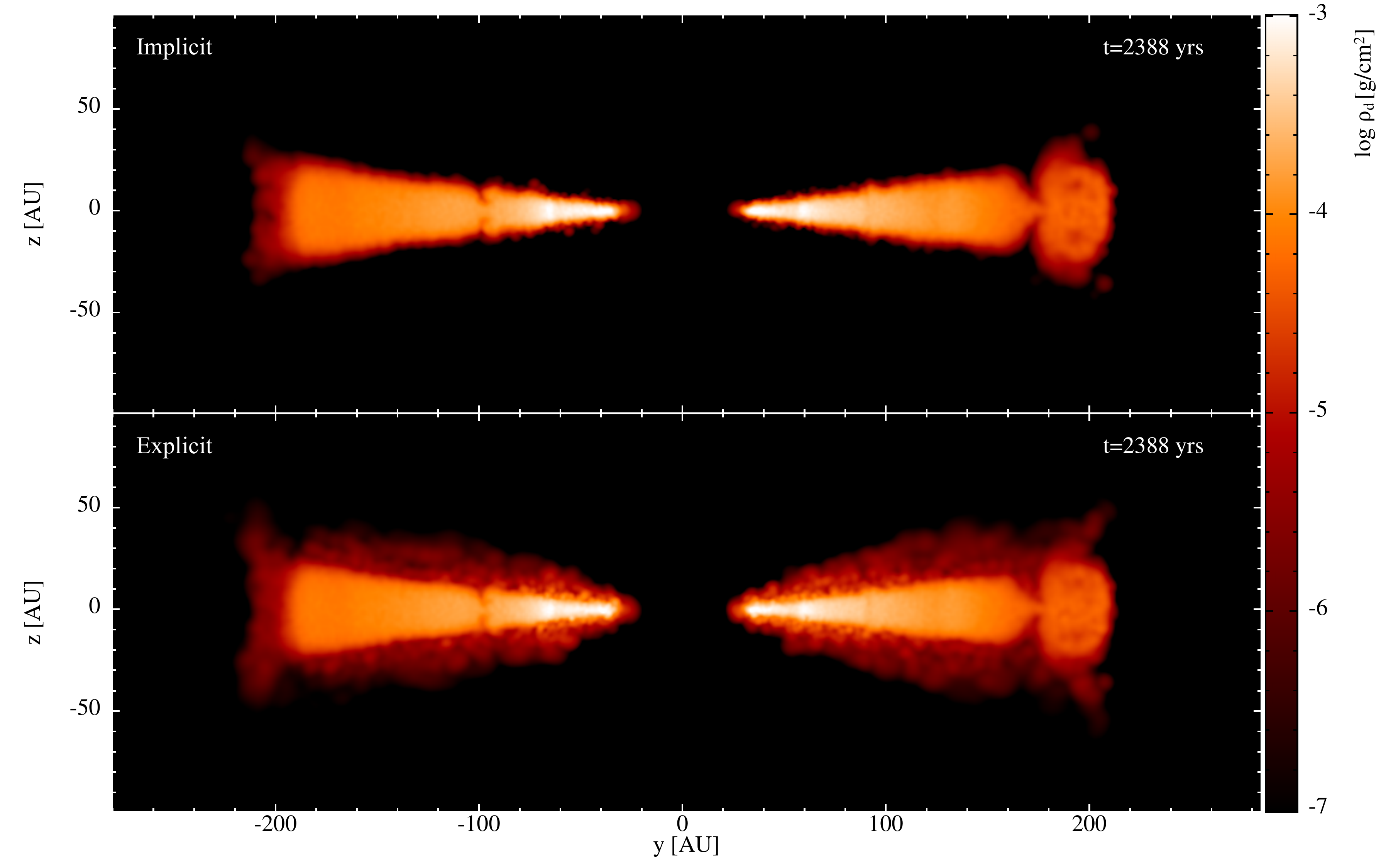}
    \caption{Cross section of the dust density in the protoplanetary disc with an embedded planet. \textit{Upper panel:} Calculation using the implicit dust algorithm. \textit{Lower panel:} Calculation using the explicit dust algorithm. In the explicit calculation there is an enhancement in dust density above (and below) the bulk of where the dust is expected to be. This is an artefact of the dust variable becoming negative.  Using the implicit method avoids this problem.}
    \label{fig:ppd_xsec}
\end{figure*}

Here we demonstrate the effect of the dust variable going negative in regions with a large dust fraction gradient also occurs in more complex simulations. From the same protoplanetary disc calculations that were discussed in Section \ref{sec:ppd} we take a cross section through both discs. This is shown in Fig. \ref{fig:ppd_xsec}, the upper panel shows the dust density cross section of the disc from the implicit calculation and the lower is the same for the explicit algorithm. Whilst the bulk of the dust across both calculations is settling towards the midplane of the disc, in the calculation that used the explicit dust algorithm there is an apparent excess of dust above and below the the bulk of the dust layer. In addition to this enhancement above and below the midplane, there is also excess dust in the gap carved by the planet. This is due to the dust variable going negative in these regions. This is the same effect as that demonstrated by the simple dust settling test but in a full 3D global simulation. Again, the implicit method for evolving the dust diffusion avoids this problem. 

\section{Conclusions}

In this paper we have described an implicit algorithm to solve the dust-as-mixture dust diffusion equation of \citet{price_small_grains_2015MNRAS.451..813P}. We have benchmarked this algorithm against standard tests in the literature. The algorithm has also been tested in the settings of the collapse of a dense molecular cloud core and in a protoplanetary disc containing a planet, and compared with results and methods in the literature. The main features of the algorithm we have described are summarised as follows
\begin{enumerate}
    \item The timestep criterion required by explicit dust diffusion can be avoided meaning that the dust evolution can be evolved on hydrodynamical timesteps which, depending on the dust size, can be much larger.
    \item The time to solution compared to a fully explicit solution can be greatly reduced in the regimes of intermediate to large dust grains (by orders of magnitude).  Even for very small dust grains, the implicit method is not much slower.
    \item Due to the above there is no need to artificially limit the stopping time of dust particles, which has the adverse effect of essentially changing the dust grain size.
    \item The iterative solution of the dust diffusion equation in the implicit method naturally avoids the dust variable becoming negative. This avoids dust apparently `leaking' through strong dust fraction gradients.
\end{enumerate}
Overall we find this implicit algorithm to be fast computationally, to solve the dust diffusion equations accurately, and to avoid several problems that manifest in explicit calculations. However, as with any dust-as-mixture method, the method still fails to give accurate solutions in the limit of large (weakly coupled) grains for which the stopping time is long.  Using a dust-as-particles method is more appropriate in such circumstances.

\section*{Acknowledgements}

The authors thank the anonymous referee for their useful comments. DE thanks Pablo Loren-Aguilar for useful discussions, and Thomas J. R. Bending for guidance regarding the \textsc{sphNG} code. DE is funded by a Science and Technology Facilities Council (STFC) studentship (ST/V506679/1). The calculations discussed in this paper were performed on the University of Exeter Supercomputer, ISCA, and on the DiRAC Data Intensive system, operated by the University of Leicester IT Services, which forms part of the STFC DiRAC HPC Facility (www.dirac.ac.uk). The latter equipment is funded by BIS National E-Infrastructure capital grant ST/K000373/1 and STFC DiRAC Operations grant ST/K0003259/1. DiRAC is part of the National e-Infrastructure. Several figures were made using \textsc{splash} \citep{price_splash_2007} and \textsc{matplotlib} \citep{Hunter:2007}.
\section*{Data Availability}

The data and scripts used to produce the figures in this article can be found at \href{https://doi.org/10.5281/zenodo.10794912}{https://doi.org/10.5281/zenodo.10794912}. A description of the files and scripts are also provided.



\bibliographystyle{mnras}
\bibliography{refs} 




\appendix

\section{Solving the quartic analytically}\label{App:qaurtic}

The solutions to the quartic equation
\begin{equation}
    P(x) = x^4 + a_3 x^3 + a_2 x^2 + a_1 x + a_0 = 0,  
\end{equation}
can be found via solving the resolvent cubic if $P(x)$ is depressed, i.e. $a_3 = 0$. Firstly, the real root, $y_1$ is found for the resolvent cubic
\begin{equation}
    y^3 - a_2 y^2 + - 4a_0y + 4 a_2 a_0 - a_1^2 = 0.
\end{equation}
Then the four solutions to the quartic $P(x)$ are given by the four solutions of \citep{abramowitz_handbook_1972}
\begin{equation}
    z^2 \mp \left( y_1 - a_2 \right)^{1/2} z + \left( \frac{y_1}{2} \mp \left( \frac{y_1}{2}\right)^2 - a_0 \right) = 0.
\end{equation}


\bsp	
\label{lastpage}
\end{document}